\documentclass[twocolumn]{aastex631}

\usepackage{cleveref}

\begin{document}

\title{AllBRICQS: The Discovery of Luminous Quasars in the Northern Hemisphere}

\author[0009-0000-2050-1118]{Yunyi Choi}
\affiliation{SNU Astronomy Research Center, Department of Physics \& Astronomy, Seoul National University, \\1 Gwanak-ro, Gwanak-gu, Seoul 08826, Republic of Korea
}

\author[0000-0002-0759-0504]{Yuming Fu}
\affiliation{Leiden Observatory, Leiden University, Einsteinweg 55, 2333 CC Leiden, The Netherlands}
\affiliation{Kapteyn Astronomical Institute, University of Groningen, P.O. Box 800, 9700 AV Groningen, The Netherlands}

\author[0000-0002-8537-6714]{Myungshin Im}
\affiliation{SNU Astronomy Research Center, Department of Physics \& Astronomy, Seoul National University, \\1 Gwanak-ro, Gwanak-gu, Seoul 08826, Republic of Korea
}

\author[0000-0002-7350-6913]{Xue-Bing Wu}  
\affiliation{Department of Astronomy, School of Physics, Peking University, Beijing 100871, People's  Republic of China}
\affiliation{Kavli Institute for Astronomy and Astrophysics, Peking University, Beijing 100871, People's  Republic of China}

\author[0000-0003-0017-349X]{Christopher A. Onken}
\affiliation{Research School of Astronomy and Astrophysics, Australian National University, Cotter Road Weston Creek, ACT 2611, Australia}
\affiliation{Centre for Gravitational Astrophysics, Australian National University, Building 38 Science Road, Acton, ACT 2601, Australia}

\author[0000-0002-4569-016X]{Christian Wolf}
\affiliation{Research School of Astronomy and Astrophysics, Australian National University, Cotter Road Weston Creek, ACT 2611, Australia}
\affiliation{Centre for Gravitational Astrophysics, Australian National University, Building 38 Science Road, Acton, ACT 2601, Australia}

\author[0000-0002-3118-8275]{Seo-Won Chang}  
\affiliation{SNU Astronomy Research Center, Department of Physics \& Astronomy, Seoul National University, \\1 Gwanak-ro, Gwanak-gu, Seoul 08826, Republic of Korea
}

\author[0000-0003-4422-6426]{Hyeonho Choi}  
\affiliation{SNU Astronomy Research Center, Department of Physics \& Astronomy, Seoul National University, \\1 Gwanak-ro, Gwanak-gu, Seoul 08826, Republic of Korea
}

\author[0009-0003-1280-0099]{Mankeun Jeong}  
\affiliation{SNU Astronomy Research Center, Department of Physics \& Astronomy, Seoul National University, \\1 Gwanak-ro, Gwanak-gu, Seoul 08826, Republic of Korea
}

\author[0000-0003-1647-3286]{Yongjung Kim} 
\affiliation{School of Liberal Studies, Sejong University, 209 Neungdong-ro, Gwangjin-Gu, Seoul 05006, Republic of Korea}

\author[0000-0002-5760-8186]{Gu Lim}
\affiliation{Department of Earth Sciences, Pusan National University, Busan 46241, Republic of Korea}
\affiliation{Institute for Future Earth (IFE), Pusan National University, Busan 46241, Republic of Korea}

\author[0009-0005-3823-9302]{Yuxuan Pang}
\affiliation{Department of Astronomy, School of Physics, Peking University, Beijing 100871, People's  Republic of China}
\affiliation{Kavli Institute for Astronomy and Astrophysics, Peking University, Beijing 100871, People's  Republic of China}

\author[0009-0001-3758-9440]{Taewan Kim}
\affiliation{SNU Astronomy Research Center, Department of Physics \& Astronomy, Seoul National University, \\1 Gwanak-ro, Gwanak-gu, Seoul 08826, Republic of Korea
}

\author[0000-0002-9254-144X]{Jubee Sohn}
\affiliation{SNU Astronomy Research Center, Department of Physics \& Astronomy, Seoul National University, \\1 Gwanak-ro, Gwanak-gu, Seoul 08826, Republic of Korea
}

\author[0000-0002-6925-4821]{Dohyeong Kim}
\affiliation{Department of Earth Sciences, Pusan National University, Busan 46241, Republic of Korea}

\author[0000-0002-1418-3309]{Ji Hoon Kim}
\affiliation{SNU Astronomy Research Center, Department of Physics \& Astronomy, Seoul National University, \\1 Gwanak-ro, Gwanak-gu, Seoul 08826, Republic of Korea
}

\author[0000-0002-8130-8044]{Eunhee Ko}
\affiliation{SNU Astronomy Research Center, Department of Physics \& Astronomy, Seoul National University, \\1 Gwanak-ro, Gwanak-gu, Seoul 08826, Republic of Korea
}

\author[0000-0002-6639-6533]{Gregory S. H. Paek}  
\affiliation{Institute for Astronomy, University of Hawaii, 2680 Woodlawn Drive, Honolulu, HI 96822, USA}

\author[0009-0009-5870-4266]{Sungho Jung}
\affiliation{SNU Astronomy Research Center, Department of Physics \& Astronomy, Seoul National University, \\1 Gwanak-ro, Gwanak-gu, Seoul 08826, Republic of Korea
}

\correspondingauthor{Myungshin Im}
\email{myungshin.im@gmail.com}

\begin{abstract}
We present the second catalog of bright quasars from the All-sky BRIght, Complete Quasar Survey (AllBRICQS), focusing on spectroscopically observed quasars in the Northern Hemisphere with Galactic latitude $|b| > 10^\circ$. This catalog includes their spectral data, redshifts, and luminosities. AllBRICQS aims to identify the last remaining optically bright quasars using data from the Wide-field Infrared Survey Explorer (WISE) and Gaia all-sky survey Data Release 3 (DR3). AllBRICQS searches for quasars that are brighter than $B_P = 16.5$ or $R_P = 16$~mag in Gaia DR3, based on simple selection criteria. Here, we report 62 new AllBRICQS quasars spanning various types, which include typical broad emission line quasars and the most luminous iron low-ionization broad absorption line quasars discovered to date. Spectroscopic observations were conducted using the Long-Slit Spectrograph on the 1.8-meter telescope at Bohyunsan Optical Astronomy Observatory, YFOSC on the 2.4-meter telescope at Lijiang Observatory, and BFOSC on the 2.16-meter telescope at Xinglong Observatory. We applied flux calibration using ZTF broadband photometry to correct for attenuation due to intermittent thin clouds during the observations. Redshifts were determined using inverse-variance weighted cross-correlation methods. Our targets span the bolometric luminosity range of $44.9<\log \left( L_{\rm bol} / {\rm erg~s^{-1}} \right)<48.0$ at redshifts between 0.09 and 2.48. These confirmed AllBRICQS quasars provide a valuable resource for future research into quasar evolution, black holes, their environments, and their host galaxies across multiple wavelengths.

\end{abstract}

\keywords{Quasars (1319) --- Active galactic nuclei (16) --- Spectroscopy (1558) --- Redshift surveys (1378) --- Astronomical catalogs (205)}

\section{Introduction} \label{sec:intro}

A quasar is an active galactic nucleus (AGN) with a bolometric luminosity of $L_{\mathrm{bol}} \geq 10^{45}$~erg~s$^{-1}$. Powered by a central black hole (BH), a quasar interacts with its host galaxy and influences the surrounding circumgalactic medium through energetic outflows. When a quasar is optically bright, its high signal-to-noise ratio (SNR) facilitates detailed investigations into its physical mechanisms and interactions with the environment.

From more common, modestly growing BHs to rare, rapidly growing BHs at greater distances, bright quasars provide valuable insights into the mass and Eddington ratio of their central BHs. Since a brighter quasar flux is associated with a larger broad line region (BLR) \citep{bentz_low-luminosity_2013}, direct high spatial resolution investigation can enable direct imaging of the BLR, allowing precise measurement of its size \citep{gravity_collaboration_spatially_2018, gravity_collaboration_spatially_2020, gravity_collaboration_central_2021, gravity_collaboration_geometric_2021}. The inherent brightness of these quasars makes them ideal targets for continuous time-series observations, facilitating variability studies that deepen our understanding of accretion disk processes in active galactic nuclei (AGNs) \citep{li_ensemble_2018, sanchez-saez_quest-silla_2018}.

Because of their importance, bright quasars ($V \lesssim 17$ mag) have been searched extensively, dating back to as early as 1983 \citep{schmidt_quasar_1983, hagen_hamburg_1995, hewett_large_1995, becker_first_2001, schneider_sloan_2005, im_seoul_2007, lee_seoul_2008, yang_desi_2023, kim_exploring_2024, nakoneczny_qzo_2025}. Despite these efforts, there seems to be a substantial number of bright quasars that have been missed in previous surveys. Recent discoveries of new bright quasars demonstrates that previous studies have missed such objects \citep{lee_seoul_2008, onken_ultraluminous_2022, wolf_accretion_2024}. The reasons for these missed detections may include low survey completeness, incomplete selection criteria, and photometric flag issues, although it remains difficult to verify conclusively. Additionally, a large number of low Galactic latitude quasars have been found recently \citep{im_seoul_2007, fu_finding_2022, werk_plane_2024, huo_finding_2025}.

More recently, the low resolution prism spectra from Gaia mission have been used to construct a comprehensive quasar candidate catalog, Quaia \citep{storey-fisher_quaia_2024}. However, this catalog may fail to identify some bright quasars, and the redshift estimates can be imprecise due to the limited spectral resolution (see Section~\ref{sec:quaia} for details). To accurately characterize quasars and derive key physical properties such as black hole mass, higher-resolution spectroscopic observations are essential.

To address the need for identifying the remaining optically bright quasars, the All-sky BRIght, Complete Quasar Survey (AllBRICQS) project \citep{onken_allbricqs_2023} was initiated as an ongoing effort to discover previously undetected bright quasars using all-sky data: Gaia Data Release 3 (DR3) \citep{gaia_collaboration_gaia_2016, gaia_collaboration_gaia_2021} and the Wide-field Infrared Survey Explorer (WISE) \citep{wright_wide-field_2010, mainzer_initial_2014}. The first publication in this series, \citet{onken_allbricqs_2023}, reported 156 spectroscopically confirmed quasars discovered using the Wide Field Spectrograph \citep[WiFeS; ][]{dopita_wide_2007, dopita_wide_2010} on the Australian National University (ANU) 2.3-m telescope at Siding Spring Observatory (SSO). This selection method, based on four straightforward criteria (detailed in Section~\ref{sec:cansel}), demonstrated high efficiency in identifying bright quasars. Following on this effort, we have conducted spectroscopic observations of ALLBRICS quasar candidates in the northern hemisphere.

This paper presents the results from the spectroscopic observations of AllBRICQS quasar candidates in the Northern Hemisphere. Section~\ref{sec:cansel} outlines the quasar candidate selection criteria established in the previous AllBRICQS paper \citep{onken_allbricqs_2023}, discussing the purity and completeness of this selection method. Section~\ref{sec:obs} describes the observations carried out using telescopes at three different observatories. Section~\ref{sec:datacal} details the spectral data reduction process, including flux scaling, target classification, redshift estimation, and bolometric luminosity measurements. Section~\ref{sec:allsam} presents the classification results of the AllBRICQS sample, discusses their properties, and highlights quasars with distinctive spectral features. Section~\ref{sec:discus} provides a comparison between our newly discovered quasars and those from other surveys. Finally, Section~\ref{sec:conclu} summarizes the main findings of this study.

We adopt a flat $\Lambda$CDM cosmology with a matter density of $\Omega_{\mathrm{m}}=0.3$ and a Hubble-Lemaître constant of $H_0 = 70$~km~s$^{-1}$~Mpc$^{-1}$. Throughout this paper, all magnitudes are given in the Vega system unless otherwise specified.

\section{Candidate Selection} \label{sec:cansel}

AllBRICQS \citep{onken_allbricqs_2023} identifies quasar candidates more effectively than previous methods by utilizing data from the WISE catalogue \citep{wright_wide-field_2010, mainzer_initial_2014} and precise parallax and proper motion measurements from Gaia DR3. The AllBRICQS selection method focuses on quasar candidates with $B_P < 16.5$ or $R_P < 16.0$~mag, a range where stars, galaxies, and quasars can be more reliably distinguished. The selection criteria, thoroughly detailed in \citet{onken_allbricqs_2023}, consist of four key requirements:

\begin{enumerate}
    \item Galactic Latitude $|b| > 10^\circ$: Reduces contamination from the Galactic plane where stellar density is high.
    \item Gaia Parallax and Proper Motion (PPM) $< 4\sigma$: Ensures the selection of extragalactic sources by filtering out objects with significant measurable motion, typical of stars within the Milky Way.
    \item $B_P / R_P$ Excess Factor $< 1.5$ (Point Source)\footnote{This selection may exclude AGNs with significant host galaxy contributions detectable by Gaia DR3, thereby introducing a bias against AGNs or quasars at $z < 0.3$, as well as against lensed quasars with resolved lens galaxies or blended lensed images.}: Differentiates between resolved (galaxies) and unresolved (stars and quasars) sources. It refers \texttt{phot\_bp\_rp\_excess\_factor} \citep[C;][]{evans_gaia_2018, riello_gaia_2021}.
    \item WISE $W1-W2 > 0.2$~mag: Utilizes mid-infrared color selection to distinguish AGNs from stars and low-redshift galaxies. Profile-fit magnitudes from the AllWISE catalog\footnote{\url{https://wise2.ipac.caltech.edu/docs/release/allwise/expsup/index.html}} are used.
\end{enumerate}

Although the original AllBRICQS selection process inadvertently used AllWISE magnitudes, it was intended to rely on CatWISE photometry \citep{eisenhardt_catwise_2020, marocco_catwise2020_2021}, which offers improved source deblending and astrometric precision. In this work, we present all $W1-W2$ color values using CatWISE magnitudes when available, and supplement with AllWISE when necessary. This correction affects only a small number of sources near the $W1-W2 = 0.2$ threshold and has minimal impact on the overall sample distribution.

By applying these criteria, the first AllBRICQS paper \citep{onken_allbricqs_2023} identified 290 quasar candidates. Among them, 166 were observed using the ANU 2.3-meter telescope at SSO with the WiFeS instrument \citep{dopita_wide_2007,dopita_wide_2010}. Spectroscopic confirmation was achieved for 156 candidates, including 140 newly discovered quasars. Based on these results, \citet{onken_allbricqs_2023} estimated the selection purity of the AllBRICQS method to be approximately 96\%. Many of the remaining unobserved candidates have since been followed up at facilities in the Northern Hemisphere.

To evaluate the completeness of our selection method, \cite{onken_allbricqs_2023} tested it against the Million Quasars Catalogue v7.2 \citep[Milliquas;][]{flesch_million_2021} using Gaia sources with $B_P < 17$~mag, redshifts $z > 0.3$, and galactic latitudes $|b| > 10^\circ$. The completeness is approximately 96.5\%, demonstrating the robustness of our criteria. While a few quasars were excluded due to the absence of PPM measurements, the vast majority were consistent with being drawn from the statistical tails of a zero-PPM population. This underscores the stringent nature of the AllBRICQS selection, which was designed to maintain both high purity and completeness.

\section{Observations} \label{sec:obs}

\begin{figure}[t]
    \centering
    \includegraphics[width=0.5\textwidth]{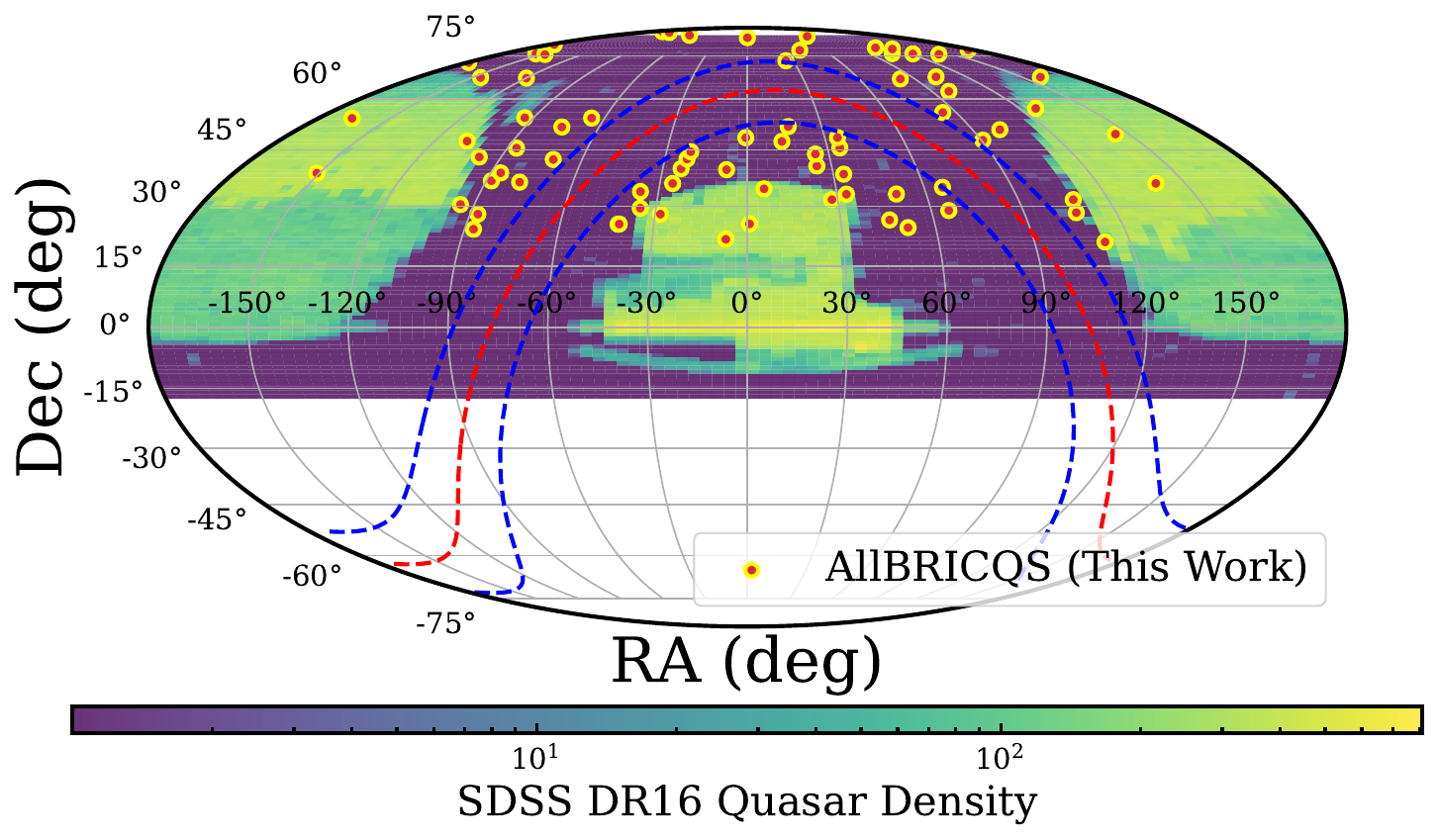}
    \caption{The sky positions of the 75 observed targets are overplotted on the SDSS footprint. Red points with yellow outlines indicate the observed sources. The red and blue dashed lines represent Galactic latitudes of $b = 0^\circ$ and $b = \pm10^\circ$, respectively.}
    \label{fig:sdssfootprint}
\end{figure}

To expand the AllBRICQS sample and investigate bright quasar candidates in the Northern Hemisphere, we conducted long-slit spectroscopic observations of 75 targets from the 82 remaining candidates yet to be observed in this region. As illustrated in Figure~\ref{fig:sdssfootprint}, these candidates are not confined to previously unexplored regions; they also include objects located within the footprint of the Sloan Digital Sky Survey Data Release 16 Quasar Catalog \citep[SDSS DR16Q;][]{ahumada_16th_2020}, as discussed in detail in Section~\ref{sec:sdss}. 

Among the 75 targets, 68 were observed with the 1.8-meter telescope at the Bohyunsan Optical Astronomy Observatory (BOAO) in South Korea. An additional 8 and 4 targets were observed with the Lijiang 2.4-meter telescope (LJT) at the Lijiang Observatory (LJO) \citep{wang_lijiang_2019} and the Xinglong 2.16-meter telescope (XLT) at the Xinglong Observatory (XLO) \citep{fan_xinglong_2016}, respectively, both located in China. Notably, five candidates were observed at multiple observatories to ensure consistency and data quality. The observations were conducted between UT 2022~October~27 and 2024~December~1.

BOAO is located at $36^\circ09'52.8''$N, $128^\circ58'35.8''$E in South Korea. Observations at BOAO utilized the Long-Slit Spectrograph, configured with a slit width of $3.6\arcsec$ and covering a wavelength range from 3500 to $10,500~\mathrm{\AA}$. The 150V grating provided a dispersion of $5.2~\mathrm{\AA}/\mathrm{pix}$, and a spectral resolution of $R \sim 280$. This spectroscopic setup closely resembles that used in the SNUQSO bright quasar survey conducted at BOAO with the same instrument \citep{im_seoul_2007, lee_seoul_2008}. Spectra were obtained over six nights between UT 2022~October~27 and 2024~December~1, covering a broad wavelength range essential for redshift determination and luminosity measurements.

The LJO is located at $26^\circ41^\prime42^{\prime\prime}\mathrm{N}, 100^\circ01^\prime48^{\prime\prime}\mathrm{E}$, while the XLO is situated at $40^\circ23^\prime39^{\prime\prime}\mathrm{N}, 117^\circ34^\prime30^{\prime\prime}\mathrm{E}$, both in China. At LJO, we used the Yunnan Faint Object Spectrograph and Camera (YFOSC), configured with slit widths of $1.8\arcsec$ and $2.5\arcsec$, covering a wavelength range of 5120–$9850~\mathrm{\AA}$. The grism provided a dispersion of $1.5~\mathrm{\AA}/\mathrm{pix}$, corresponding to a spectral resolution of $R \sim 600$ and 800 at 6600\,\AA, depending on the slit width. These spectra were acquired over four nights between UT 2022 December 13 and 2023 February 24, contributing valuable data to the AllBRICQS sample. 

At XLO, we employed the Beijing Faint Object Spectrograph and Camera \citep[BFOSC;][]{zhao_investigating_2018}, which covers a wavelength range of 4107–$6724~\mathrm{\AA}$ with a slit width of $2.3\arcsec$ and a dispersion of $1.34~\mathrm{\AA}/\mathrm{pix}$, corresponding to $R \sim 550$ at 5100\,\AA. The spectra at XLO were obtained over a single night on UT 2022 November 12.

To ensure accurate spectral calibration, we observed various spectrophotometric standard stars alongside the AllBRICQS candidates. These standard stars were used to calibrate the overall spectral shape. However, since the standard stars were not necessarily observed under identical conditions, we also applied photometry data for absolute flux calibration. Instead, flux scaling was addressed separately, as detailed in Section~\ref{sec:flscl}.

\section{Data Calibration} \label{sec:datacal}
\subsection{Data Reduction} \label{sec:redc}
For BOAO observational data, we employed the Image Reduction and Analysis Facility \citep[IRAF;][]{tody_iraf_1986, tody_iraf_1993} for standard reduction procedures, including bias subtraction and flat-field correction. Cosmic ray removal was performed using \texttt{astroscrappy}, an optimized implementation of the \citet{van_dokkum_cosmic-ray_2001} algorithm within the \texttt{astropy} framework \citep{mccully_astropyastroscrappy_2018}.

Wavelength calibration for the BOAO data was performed by matching observed emission lines to a reference spectrum from a FeNeArHe arc lamp. This lamp provides a wide range of well-defined emission lines, enabling accurate calibration across the full spectral range. The resulting wavelength calibration error was consistently smaller than the spectral dispersion of $5.2~\mathrm{\AA}/\mathrm{pix}$, confirming the reliability of the calibration procedure. In addition, heliocentric velocity corrections were applied based on the location and timing of each observation to account for Earth's motion relative to the observed targets.

For initial flux calibration, we utilized spectrophotometric standard stars observed on the same night, at similar airmass, and within an intermediate time frame relative to the science targets. The reference spectra of the standard stars were derived from theoretical models and observational data from STIS, MMT, and ESO \citep{hamuy_southern_1992, hamuy_southern_1994, bohlin_techniques_2014, bohlin_new_2020}.

The reliable wavelength range of each spectrum varied slightly depending on the observation date. Notably, data from UT 2022 October 27 exhibited a problem with focus at longer wavelengths, leading to unreliable flux measurements (S/N~$< 5$) at $\lambda \gtrsim 7600\AA$. Despite this anomaly, spectra from the other BOAO observing dates provide reliable flux measurements over the 4000--9000\,\AA\ range. These calibrated and reduced spectra serve as the foundation for further analysis, including redshift and bolometric luminosity estimations detailed in the following sections. 

The spectroscopic data from the LJT/YFOSC and XLT/BFOSC were reduced using a Python package \texttt{pyfosc} \citep{yuming_fu_2024_10967240}, which follows the standard procedures of IRAF \citep{tody_iraf_1986, tody_iraf_1993} for spectroscopic data reduction. The cosmic ray removal was also conducted with \texttt{astroscrappy} \citep{van_dokkum_cosmic-ray_2001, mccully_astropyastroscrappy_2018}. A FeAr arc lamp was used for wavelength calibration of the XLT/BFOSC data, while a HeNeAr arc lamp was used for the LJT/YFOSC data. The reliable wavelength range of each spectrum is 5100 -- 9300 \AA\ for the LJT/YFOSC data and 4100 -- 6700 \AA\ for the XLT/BFOSC data. 

\begin{figure*}[t]
    \centering
    \includegraphics[width=\textwidth]{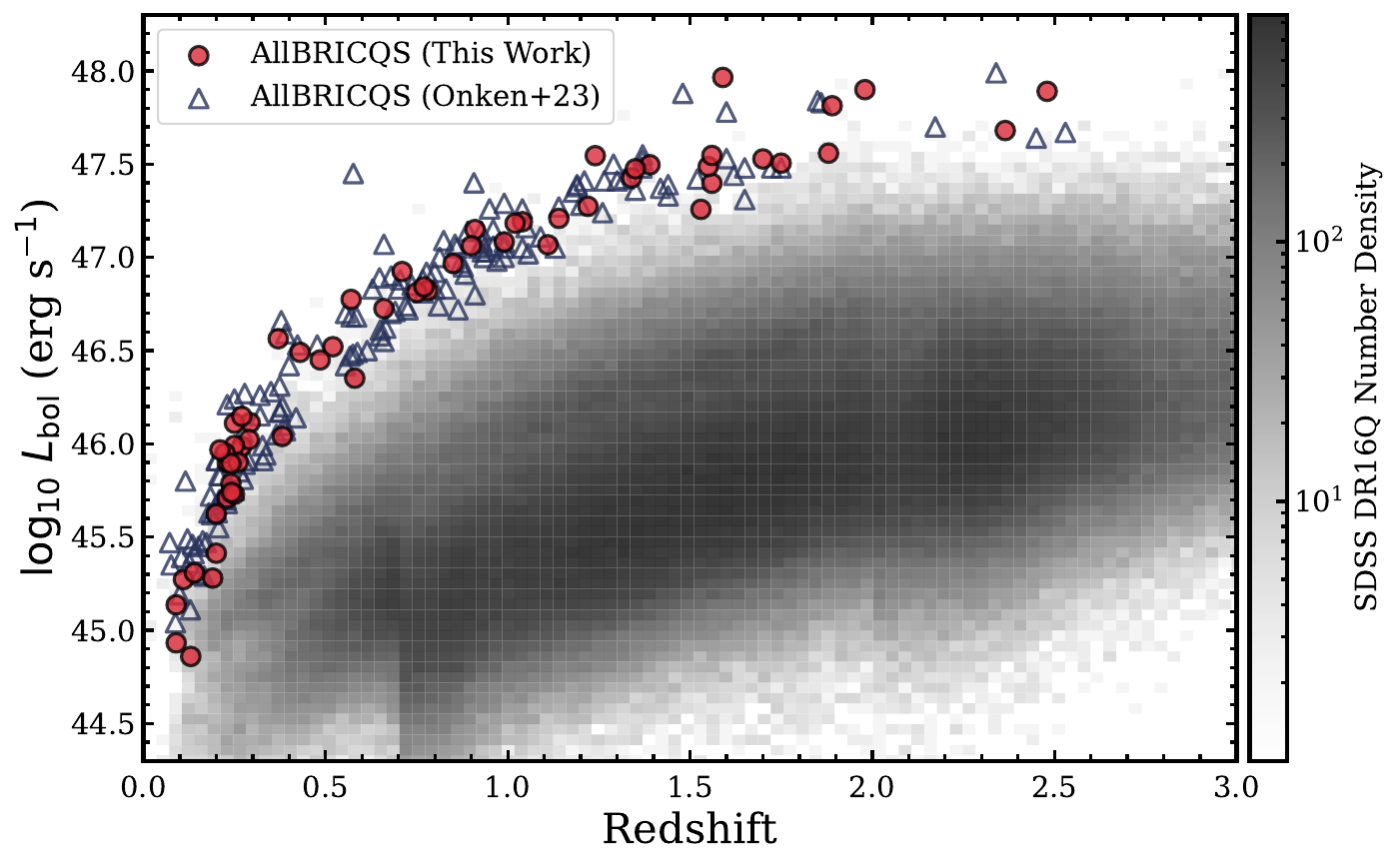}
    \caption{Distribution of redshift and bolometric luminosity for AllBRICQS quasars in this work (Northern Hemisphere; red circles) and from \citet{onken_allbricqs_2023} (Southern Hemisphere; blue triangles), in comparison to the SDSS DR16 Quasar Catalog \citep[SDSS DR16Q;][]{lyke_sloan_2020}, shown as grey contours. The AllBRICQS quasars occupy the bright end of the quasar luminosity distribution, exhibiting significantly higher luminosities than the bulk of the SDSS DR16Q sample. The grey contours correspond to a two-dimensional histogram constructed using 10,000 bins (100 bins per axis), effectively illustrating the density distribution of the SDSS quasars.}
    \label{fig:fig2}
\end{figure*}

\subsection{Flux Scaling} \label{sec:flscl}
Accurate flux scaling is essential to ensure that spectroscopic data reflect the real-time flux levels of quasar targets. Due to intermittent thin cloud coverage during our observing run, some target fluxes were attenuated. To correct for this, we applied an additional flux scaling step using light curve data from the Zwicky Transient Facility Data Release 22 \citep[ZTF DR22;][]{masci_zwicky_2019, bellm_zwicky_2019}, following the initial spectrophotometric calibration with standard stars. The flux scaling process utilizes $r$-filter photometry acquired on the exact date of our spectroscopic observations, providing a robust match to the spectral data and mitigating the effects of both atmospheric transparency variations and intrinsic quasar variability.

For the XLT data, where the spectral coverage does not fully overlap with the $r$-filter transmission range, we used $g$-filter photometry instead. If photometric data were unavailable for the observation date, we employed linear interpolation of the ZTF light curve to estimate the expected magnitude. For data collected on UT 2024 December 1, since the ZTF DR22 dataset includes observations only up to UT 2024 July 1, we flux-scaled the spectra using the most recent available photometric data to minimize the impact of temporal discrepancies. The synthetic photometry results for the ZTF $g$- and $r$-filters of the quasars are presented in Appendix ~\ref{sec:ZTFmagQuasar}.


\begin{figure*}[t]
    \centering
    \includegraphics[width=\textwidth, keepaspectratio]{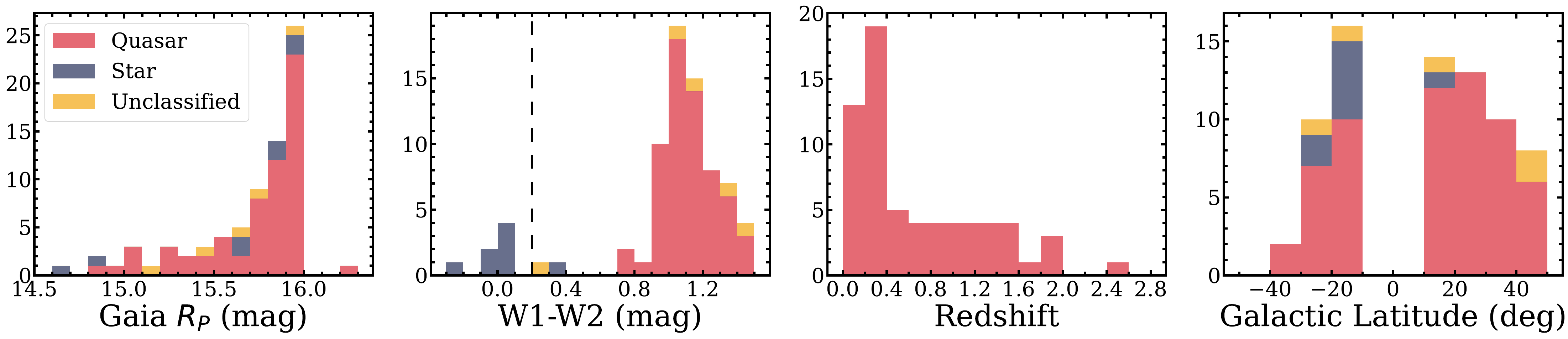}
    \caption{   
    Distributions of key properties of the spectroscopically observed 75 sources in this study. The histograms, displayed from left to right, show Gaia $R_P$ magnitude, $W1-W2$ color, measured redshift, and Galactic latitude. The $W1-W2$ color cut at 0.2 is indicated by a black dashed line. The color values are primarily sourced from the CatWISE catalog \citep{eisenhardt_catwise_2020, marocco_catwise2020_2021} and supplemented with AllWISE when CatWISE values are unavailable. Although the original AllBRICQS selection used AllWISE magnitudes, the color values shown here are based on CatWISE to reflect the intended selection methodology. As a result, a few objects appear below the 0.2 cut, since they were selected using AllWISE but would not have satisfied the cut under CatWISE photometry. With the exception of redshift, all other properties are derived from Gaia DR3 \citep{gaia_collaboration_gaia_2016, gaia_collaboration_gaia_2021}.}
    \label{fig:fig1}
\end{figure*}

\subsection{Target Classification and Redshift Measurement} \label{sec:redshift}

Spectral classification and redshift estimation were initially conducted through visual inspection of the observed spectra. In cases where classification was ambiguous, the MARZ online application was used to assist the identification process \citep{hinton_marz_2016, hinton_marz_2016-1}. 

Final redshift values (heliocentric redshifts) and their associated uncertainties were derived using \texttt{RVSNUpy} \citep{kim_rvsnupy_2025}, a Python-based tool that estimates redshifts by performing inverse-variance weighted cross-correlation between the observed spectra and a library of rest-frame template spectra. Approximately 79\% of the quasar sample exhibits redshift uncertainties ($\delta z$) smaller than 0.0005, while 92\% have $\delta z < 0.001$. For a small number of quasars, redshift values and redshift uncertainties could not be determined due to either blended emission and absorption features or low signal-to-noise ratio (SNR) in regions containing key spectral lines. 

In Table~\ref{tab:AllCofQuasar}, the redshift measurements are reported in the “AllBRICQS” column under the “Redshift” header. For targets whose redshifts could not be determined using \texttt{RVSNUpy}, approximate values with reduced significant figures and large uncertainties are listed. Note that additional redshift uncertainty, up to 0.004, may arise from the wavelength calibration process. This calibration-related uncertainty is not included in the redshift errors presented in the table.

\subsection{Bolometric Luminosity} \label{sec:bollum}

The bolometric luminosity is a critical parameter for characterizing quasars, providing insights into the energy output of the accretion process around BHs. We calculated the bolometric luminosity for each flux-scaled quasar spectrum, applying robust extinction corrections to ensure accurate luminosity estimates.

We used \texttt{PyQSOFit} \citep{guo_pyqsofit_2018, shen_sloan_2019} to derive continuum luminosities. For Galactic extinction correction, we applied the default settings: the dust map from \citet{schlegel_maps_1998}, a standard Milky Way reddening parameter of $R_V = 3.1$, and the extinction model from \citet{fitzpatrick_correcting_1999}. The continuum fitting was performed over the 4000--9000\,\AA\ range, using spectral regions selected to be relatively free of strong emission lines. The continuum model consists of a power-law component combined with a third-order polynomial to account for broad spectral curvature. Bolometric luminosities were then estimated using the method outlined in \citet{runnoe_erratum_2012, runnoe_updating_2012}, which provides empirically calibrated bolometric corrections at rest-frame wavelengths of 1450\,\AA, 3000\,\AA, and 5100\,\AA. When multiple continuum luminosities were available, we adopted the mean value (see Table \ref{tab:QuasarBLum} of Appendix~\ref{sec:LumQuasar}).

To account for host galaxy dust extinction, we performed SED fitting using photometric magnitudes from SDSS, Pan-STARRS, 2MASS, and WISE, incorporating a reddening law from \citet{fitzpatrick_correcting_1999} and following the method described in \citet{kim_estimators_2023}. We calculated the AGN continuum luminosities at 3.4\,$\mu$m and 4.6\,$\mu$m, and then derived the bolometric luminosities based on the relations presented by \citet{kim_estimators_2023, kim_eddington_2024-1, kim_eddington_2024}, applying extinction corrections. The resulting bolometric luminosities from the extinction-corrected infrared continuum are in reasonable agreement with those estimated from the rest-frame ultraviolet and optical continuum luminosities at 1450\,\AA, 3000\,\AA, and 5100\,\AA. This agreement suggests that our sample does not include any heavily dust-obscured quasars.

Figure~\ref{fig:fig2} displays the measured redshift and $\log$ bolometric luminosity values of AllBRICQS quasars. These quasars predominantly occupy the bright end of the quasar distribution, represented by the grey contours of SDSS DR16 quasars \citep{lyke_sloan_2020}. Note that J0633+6225 and J1809+2758, observed with XLT and LJT respectively, are excluded from the figure. Their exclusion is due to insufficient wavelength coverage, which is essential for accurate bolometric luminosity estimation. The redshifts of these two targets are 0.33 and 2.41, respectively.

\section{AllBRICQS Samples} \label{sec:allsam}

This section presents the classification results of the 75 observed spectra, including confirmed quasars, unclassified sources, and stars. The properties of the observed targets are summarized in Figure~\ref{fig:fig1}. A complete list and spectral gallery of confirmed quasars are provided in Table~\ref{tab:AllCofQuasar} and Appendix~\ref{sec:GalQuasar}, ordered by increasing right ascension (RA) and redshift, respectively. Similarly, the unclassified sources and stars are listed in Table~\ref{tab:AllUnknown} and Table~\ref{tab:AllStar}, with their corresponding spectral galleries shown in Appendix~\ref{sec:unclass} and Appendix~\ref{sec:stars}, respectively.

The targets were divided into two priority classes based on their \textit{WISE} color. Priority~1 targets are defined as sources with $W1 - W2 > 0.3$~mag, based on AllWISE profile-fit magnitudes. Priority~2 targets have $W1 - W2 \leq 0.3$~mag and lie closer to the stellar locus in color space. The selection purity of the Priority~1 sample is 98.4\%. Notably, all stellar contaminants in our sample—except for one—were initially classified as Priority~2 targets.

\subsection{Confirmed Quasars}

Among the 75 observed candidates, 62 sources have been confirmed as quasars through spectroscopic analysis. The newly discovered quasars exhibit $\log$ bolometric luminosities ranging from $\log L_{\rm bol} \sim 44.9$ to $48.0$~erg~s$^{-1}$ and redshifts spanning from 0.09 to 2.48. When we cross-matched them to the quasars in the Milliquas v8 catalog \citep{flesch_million_2023}, 19 of them were identified as quasar candidates based on optical to mid-infrared photometry \citep{secrest_identification_2015, richards_bayesian_2015, gaia_collaboration_gaia_2023-1, klemola_vizier_1994}. However, none of these 62 quasars had previously been spectroscopically confirmed or had their redshifts measured. In Table~\ref{tab:AllCofQuasar}, 62 new quasars are presented with information. 

For J0919+3557, we adopted the classification as a quasar based on the publicly available DESI spectrum\footnote{\url{https://www.legacysurvey.org/viewer/desi-spectrum/dr1/targetid2305843014536273669}}, rather than from our own observations. This object is classified as a galaxy at $z = 0.18$ in the Dark Energy Spectroscopic Instrument (DESI) DR1 catalog \citep[Juneau et al., in prep.;][]{desi_collaboration_data_2025}. Due to its point-like morphology, it was classified as a star in SDSS DR18 \citep{almeida_eighteenth_2023}, and as a quasar candidate by \citet{richards_bayesian_2015}. 


However, we determine its redshift to be $z = 2.365$, consistent with a quasar, supported by the presence of multiple foreground absorption systems. Specifically, three absorption lines at 6490, 6580, and $6600\,\mathrm{\AA}$ correspond to an Fe\,\textsc{ii} absorption system at $z = 1.769$, and a doublet at 7740 and $7770\,\mathrm{\AA}$ corresponds to Mg\,\textsc{ii} at the same redshift. Additionally, an Mg\,\textsc{ii} doublet at 9190 and $9210\,\mathrm{\AA}$ indicates a second absorption system at $z = 2.285$, which also includes weaker associated Fe\,\textsc{ii} lines. These features suggest that the continuum source must lie at $z > 2.29$.

Adopting $z = 2.365$ aligns with two observed broad emission features that may correspond to Mg\,\textsc{ii} and C\,\textsc{iii}], although notably, C\,\textsc{iv} is absent. The redshift is further constrained by the presence of multiple lower-redshift absorbers and the absence of a prominent Lyman forest, implying an upper limit consistent with this redshift. One possible interpretation is that the Ly$\alpha$ emission is present but weak, and the C\,\textsc{iv} line is intrinsically faint. This object may therefore be classified as a weak-line quasar, a subclass of quasars characterized by unusually weak high-ionisation lines. Such spectra are often attributed to a deficit in ionising radiation, potentially due to cold accretion disks \citep{laor_cold_2011}. Notably, all high-ionisation lines appear absent, while only low-ionisation lines are clearly detected.

\subsection{Unclassified Sources}

Five sources remain unclassified as they do not exhibit clear quasar emission lines or Balmer absorption features characteristic of stellar atmospheres. Despite this, some of these sources have non-zero redshift estimates listed in Table~\ref{tab:AllUnknown}, indicating a potentially high-probability redshift value.

A representative example is J1040+6711, which has a redshift estimate of $z = 0.474$ from the Quaia catalog \citep{storey-fisher_quaia_2024}. However, the absence of H$\alpha$, H$\beta$, or [O\,III] emission features at expected wavelengths challenges this classification. An alternative interpretation suggests that the feature at 4868\,\AA\ could correspond to either Mg\,\textsc{ii} or Ly$\alpha$, implying that the source might be a broad absorption line quasar (BALQSO) at $z = 0.74$ or a Lyman break galaxy at $z = 2.75$. 

Both scenarios, however, remain speculative. If the object is at $z = 0.74$, the expected H$\beta$ and [O\,III] lines at 8459\,\AA\ and 8713\,\AA\ are absent. Alternatively, if it is at $z = 2.75$, the expected Si\,\textsc{ii} and C\,\textsc{ii} absorption features at 4883\,\AA\ and 5004\,\AA\ are also not observed. Overall, the redshift values of these unclassified sources should be interpreted with caution, as the identification of their spectral features is uncertain.

\subsection{Stellar Sources}
Eight sources in the AllBRICQS sample were classified as stars and are listed in Table~\ref{tab:AllStar}. Seven of them were initially tagged as priority 2 targets (defined by $W1 - W2 < 0.3$, using AllWISE profile-fit magnitudes) during the quasar candidate classification process. Their parallax-to-error ratios ($\varpi / \sigma_{\varpi}$) exceed 1.71, indicating statistically significant parallax measurements. The minimum proper motion ($\mu$) among these stars is 0.13~$\mathrm{mas\,yr^{-1}}$. In contrast, quasars in our sample exhibit $\varpi / \sigma_{\varpi}$ values that do not exceed 1.74, with the majority being negative. The maximum proper motion observed among quasars is $0.15~\mathrm{mas\,yr^{-1}}$, and the average is $0.06~\mathrm{mas\,yr^{-1}}$.

These metrics suggest that the selection purity could have been improved by applying more stringent criteria on proper motion and parallax-to-error ratio. A revised set of selection criteria is proposed in Section~\ref{sec:conclu}. Based on these updated criteria, J0319+3309, J0422+2854, and J2357+4826 are identified as stars with high probability, due to their high proper motion values ($\mu > 0.8~\mathrm{mas\,yr^{-1}}$).

\subsection{Objects of Special Interest} \label{sec:distspec}

Identifying distinctive spectral features within the AllBRICQS sample provides valuable insights into rare quasar types and contributes to a deeper understanding of AGN diversity. Two notable examples are J1356+3840, a rare iron low-ionization broad absorption line quasar (FeLoBAL), and J0141+7307 which exhibits iron UV emission lines.

J1356+3840 is the only FeLoBAL among the 62 new AllBRICQS quasars. FeLoBALs, which constitute only $\sim$0.3\% of the quasar population \citep{trump_catalog_2006}, display absorption from both high- and low-ionization species, along with Fe\,\textsc{ii} or Fe\,\textsc{iii}. These objects are particularly valuable for studying quasar outflows and understanding quasar structure and host galaxy evolution \citep{villforth_host_2019, choi_physical_2022}. As shown in Appendix~\ref{sec:GalQuasar}, J1356+3840 features broad Fe\,\textsc{ii} absorption lines on the blue side of the Mg\,\textsc{ii} emission line. 

We confirm that J1356+3840 is the brightest FeLoBAL quasar identified to date. It has a bolometric luminosity of $\log L_{\mathrm{bol}} = 47.18~\mathrm{erg~s^{-1}}$ at a redshift of $z = 1.39$. Near-infrared spectroscopic observations of this object were conducted by Fu et al. (in preparation) in 2021 using the TripleSpec instrument on the Palomar 200-inch telescope (P200/TSpec).

In addition to J1356+3840, which exhibits broad Fe\,\textsc{ii} absorption lines, several quasars in our sample show prominent iron emission features. For example, in Appendix~\ref{sec:GalQuasar}, J0141+7307 at $z=1.59$ exhibits a distinctive overlapping Fe\,\textsc{ii} UV 62 and 63 emission line. Additionally, some spectra display a prominent UV 48 feature. A notable example is J1216+7948 at $z=1.68$, where an emission line appears at 5605~\AA\. This UV Fe\,\textsc{ii} emission suggests rich iron emission regions possibly influenced by overdense regions \citep{clowes_environments_2013, harris_evidence_2013}.

\begin{figure}[t]
    \centering
    \hspace*{-1.0cm}
    \includegraphics[width=0.5\textwidth]{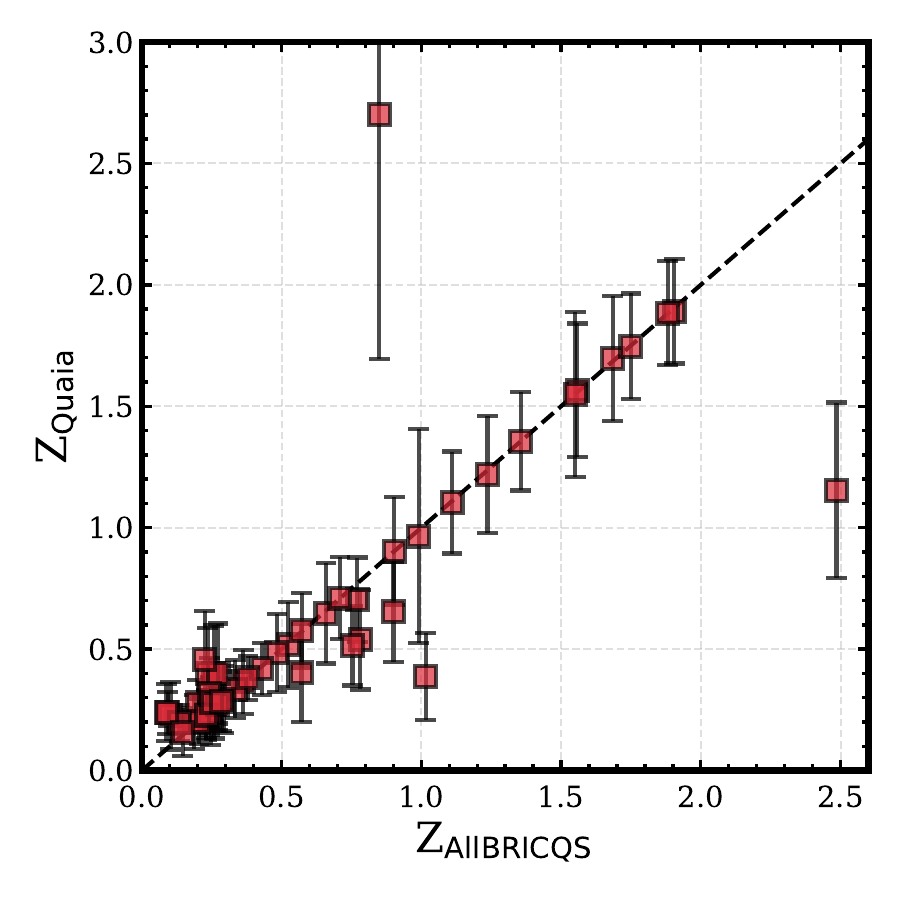}
    \caption{Comparison of redshift measurements between AllBRICQS spectroscopic values and Quaia photometric estimates. The $x$-axis represents the spectroscopic redshifts obtained from AllBRICQS observations, while the $y$-axis shows the corresponding photometric redshift estimates from the Quaia catalog.}
    \label{fig:fig3}
\end{figure}

\section{Discussion} \label{sec:discus}
\subsection{Comparison with Quaia} \label{sec:quaia}

We compared our spectroscopically confirmed AllBRICQS quasars with the $G < 20.5$ Gaia-unWISE Quasar Catalog \citep[Quaia;][]{storey-fisher_quaia_2024}. Among the 62 spectroscopically confirmed quasars, 50 are classified as quasar candidates in Quaia. This means that the completeness of the Quaia catalog is about 81\% with respect to AllBRICQS at $B_P < 16.5$ or $R_P < 16.0$~mag. None of the stars in our sample are included in the Quaia catalog, although one unclassified source (J1040+6711) is listed as a quasar candidate in Quaia.

There is a fundamental difference in selection strategy that accounts for the divergence between the two catalogs. The priority 2 candidates were designed to probe deeper into the stellar locus to test whether any quasars might exist in the tail of the color distribution. In contrast, the Quaia catalog emphasizes selection purity by relying on known quasars to train its classification model. While Quaia is optimized to recover known types of quasars, it is less likely to discover new or unusual populations. In this study, we specifically targeted objects located near the stellar locus—those that most closely resemble quasars in color space, despite being consistent with stellar colors—and spectroscopically confirmed them to be stars.

Quaia estimates redshifts using a trained $k$-nearest neighbors (KNN) model that utilizes low-resolution redshift estimates from Gaia spectra along with photometric features from Gaia and WISE, offering a purely data-driven approach to quasar classification. While the Quaia catalog provides a convenient method for estimating quasar redshifts from photometric data, discrepancies with our spectroscopic measurements were noted. As shown in Figure~\ref{fig:fig3}, the majority of discrepancies are modest, but a few significant outliers highlight the limitations of the Quaia redshifts. We find a catastrophic failure fraction of 6\% (28\%) for $\left| \Delta z / (1+z) \right| > 0.2$ (0.1). The 6\% fraction at the 0.2 threshold aligns well with the 6\% reported in \citet{storey-fisher_quaia_2024}, whereas the 28\% fraction at the 0.1 threshold is notably higher than their reported 10\%.

The most significant redshift discrepancy is observed in the case of J1410+7533, where the Quaia estimate differs by 1.85 from our measurement. As shown in Appendix \ref{sec:GalQuasar}, our spectrum places this quasar at $z = 0.85$, while Quaia estimates $z = 2.7$. This discrepancy could arise from spectral line misidentification. If the emission line around 5200\,\AA\ were identified as Si\,\textsc{iv}, a redshift of 2.7 is plausible. However, at this redshift, a prominent Ly$\alpha$ emission line should appear near 4500\,\AA. The absence of this feature leads us to reject the Quaia redshift estimate of 2.7.

The second-largest discrepancy between redshift measurements is observed for J0139+4036, with a difference of $\Delta z = 1.33$. This quasar, the most distant object in the AllBRICQS-North sample, has a spectroscopic redshift of $z = 2.48$, while Quaia estimates a significantly lower photometric redshift of $z = 1.15$. At $z = 1.15$, the Mg\,\textsc{ii} line would be expected around 6000\,\AA. However, this line is absent, and the emission features at 4800\,\AA\ and 5400\,\AA\ in the spectrum cannot be explained at this lower redshift.

\subsection{Comparison with SDSS} \label{sec:sdss}
By Data Release 8 (DR8; \citealt{aihara_eighth_2011}), SDSS had imaged over 14,000~deg$^2$ of the sky, primarily in the Northern Hemisphere. \citet{lyke_sloan_2020} provided the SDSS DR16Q catalog from the extended Baryon Oscillation Spectroscopic Survey (eBOSS), which includes spectroscopically confirmed quasars that were initially selected as candidates from imaging data. Most eBOSS quasars were selected using the \texttt{CORE} algorithm, which was applied to SDSS point sources with $g < 22$ or $r < 22$ (extinction-corrected). These candidates were then filtered using the XDQSOz algorithm \citep{bovy_photometric_2012}, requiring a probability greater than 20\% of being a quasar at $z > 0.9$, followed by a WISE–optical color cut to reduce stellar contamination.

\begin{figure}[t]
    \centering
    \hspace*{-0.5cm}
    \includegraphics[width=0.48\textwidth, keepaspectratio]{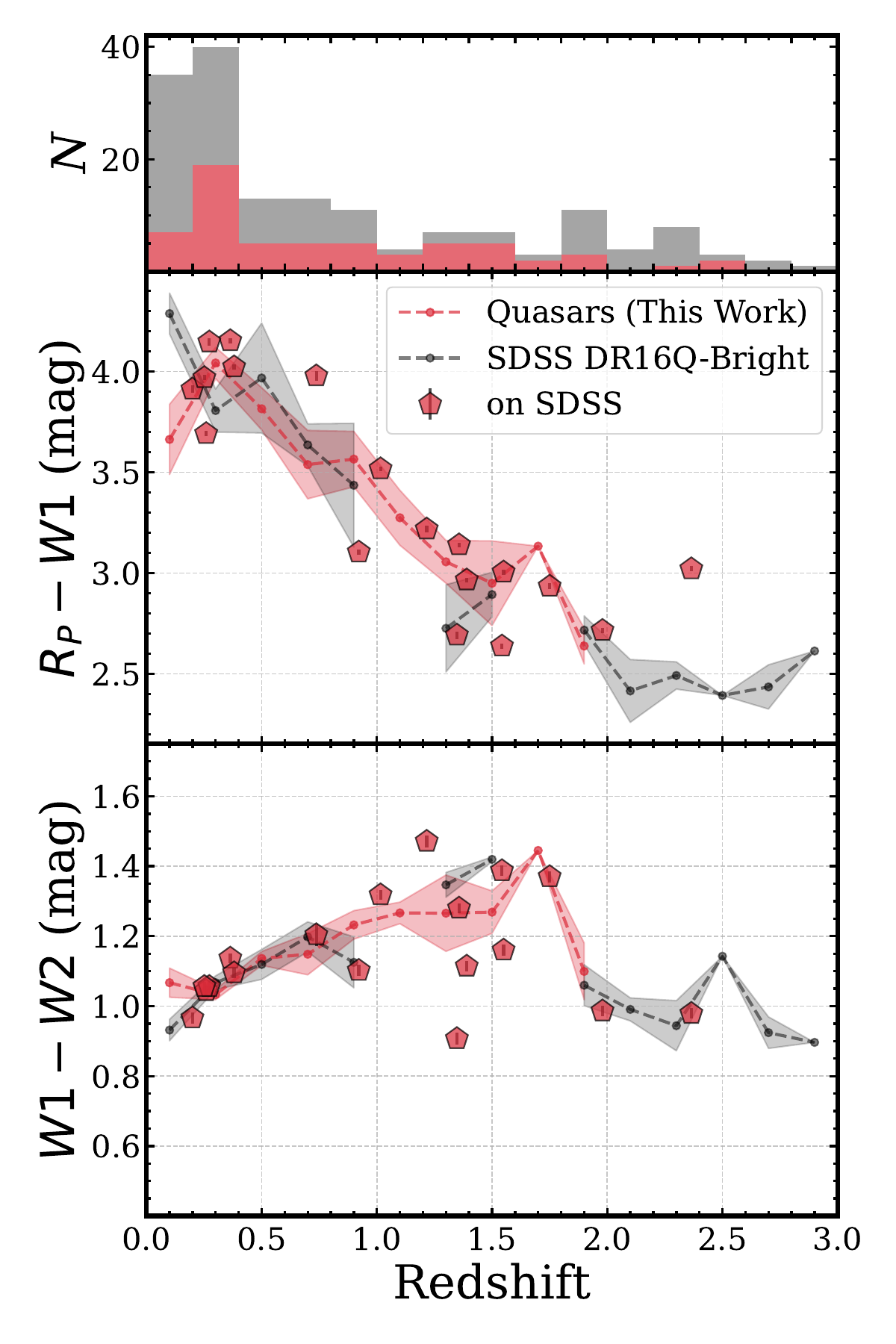}
    \caption{Distribution of $R_P - W1$ and $W1 - W2$ colors as a function of redshift. The shaded regions along the central dashed lines represent the mean color trends and the 1$\sigma$ scatter within redshift bins of width 0.2. The red shaded region corresponds to our newly discovered 62 quasars, while the black shaded region corresponds to 319 SDSS DR16Q quasars \citep{lyke_sloan_2020} that satisfy our bright selection criteria ($B_P < 16.5$ or $R_P < 16.0$~mag). Redshift bins in the SDSS sample containing only a single data point within $z < 2.0$ were excluded. Red pentagons denote the 18 quasars located within the SDSS footprint.}
    \label{fig:allbricqs_sdss}
\end{figure}

Among our 62 newly confirmed quasars, 18 lie within the SDSS footprint, but are absent\footnote{J0827+5703 is classified as a QSO and has spectroscopic data available in the SDSS SkyServer DR18.} from the SDSS DR16Q catalog. Although three of these 18 quasars are located near bright sources in the SDSS DR8 cutout images (see Appendix~\ref{sec:sdss_img}), most do not show problematic photometric flags—specifically, the \texttt{Q} value (observation quality) and the \texttt{mode} value (photometric mode) are generally acceptable. The exception is J1649+4726, which has a \texttt{Q} value of 1, indicating poor-quality photometry. Since SDSS DR16Q does not provide completeness metrics as a function of magnitude and redshift, it is difficult to determine whether the absence of these objects is statistically expected. Moreover, it is unclear at which step(s) in the SDSS quasar selection pipeline these 18 quasars were excluded. 

\begin{figure*}[t]
    \centering
    \includegraphics[width=\textwidth, keepaspectratio]{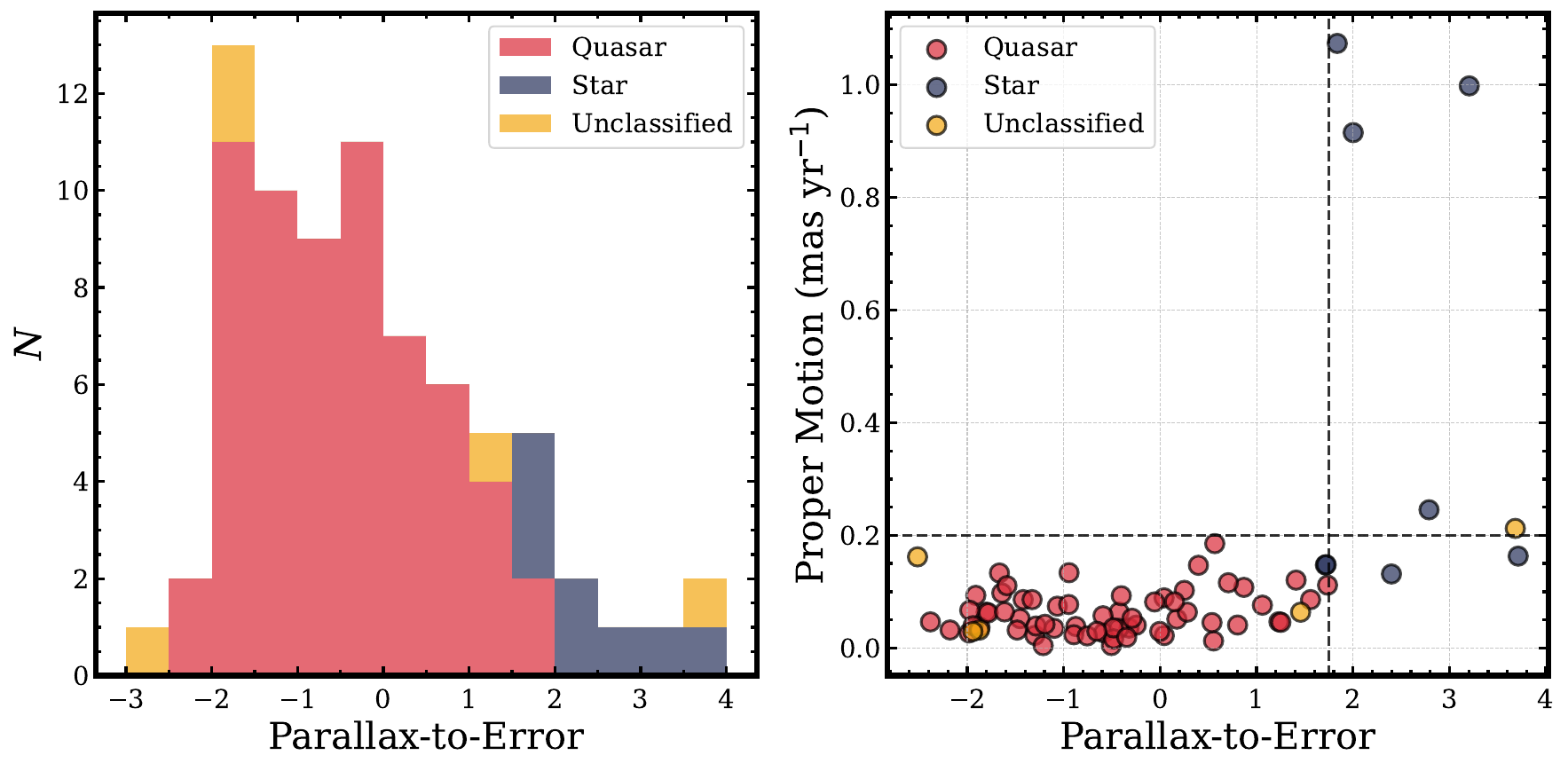}
    \caption{Distribution of the 75 observed sources in the AllBRICQS sample, with all astrometric properties derived from \textit{Gaia} DR3 \citep{gaia_collaboration_gaia_2016, gaia_collaboration_gaia_2021}. (Left) Histogram of the parallax-to-error ratio, $\varpi / \sigma_{\varpi}$. Quasars (red) do not exceed a value of 2.0, with many exhibiting negative values, which correspond to near-zero parallaxes in reality (see Section~5.2 and Figure~6 of \citealt{lindegren_gaia_2018}). Stars (navy) all have $\varpi / \sigma_{\varpi} > 1.5$. (Right) Scatter plot of proper motion $\mu$ versus parallax-to-error ratio. The dashed lines indicate the newly proposed stringent criteria for quasar selection.
    }
    \label{fig:PrlxErr_pm}
\end{figure*}

To compare the color trends of our quasars with those of SDSS DR16Q with $B_P < 16.5$ or $R_P < 16.0$~mag, we examined their color distributions as a function of redshift, as shown in Figure~\ref{fig:allbricqs_sdss}. In conclusion, no significant difference between the two distributions was observed. For both the $R_P - W1$ and $W1 - W2$ color distributions, the two datasets appear visually consistent across redshift, except in the bins $1.0-1.2$ and $1.6-1.8$, where the SDSS DR16Q sample contains only one object, making statistical interpretation unreliable.

Quantitatively, for the $R_P - W1$ color distribution shown in the upper panel of Figure~\ref{fig:allbricqs_sdss}, the energy distance measurement \citep{rizzo_energy_2016} based on 5000 permutations yields a value of 0.0703 and a $p$-value of 0.9064, indicating that the two samples are statistically consistent with being drawn from the same parent population. Similarly, for the $W1 - W2$ distribution shown in the lower panel, the energy distance is 0.189 with a $p$-value of 0.9974, further supporting the conclusion that no statistically significant difference exists between the two samples.

The pentagon points in Figure \ref{fig:allbricqs_sdss} represent quasars from our sample that are located within the SDSS footprint and therefore could have been included in the SDSS DR16Q catalog. Most of these points are found in the redshift ranges $0 -0.4$ and $1.2 -1.6$. For both redshift intervals, considering the positions of the pentagon points and the overall trends in the optical–IR and IR color distributions, a definitive reason for their omission from the SDSS catalog is difficult to identify.

\subsection{Comparison with DESI} \label{sec:desi}
The DESI Legacy Imaging Surveys Data Release 8 \citep{duncan_all-purpose_2022} covers approximately 19,000~deg$^2$, primarily in the Northern Hemisphere. DESI selects quasar candidates using photometric imaging in three optical bands ($g$, $r$, $z$) from the Legacy Imaging Surveys and two infrared bands ($W1, W2$) from WISE, applying a random forest algorithm to identify candidates in the magnitude range $16.5 < r < 23$ (AB mag) \citep{chaussidon_target_2023}.

Twenty-two of our 62 quasars lie within this footprint. The recent DESI DR1 catalog \citep[Juneau et al., in prep.;][]{desi_collaboration_data_2025} classifies 10 of our 62 quasars as confirmed quasars. Our redshift estimates are in excellent agreement with those in DESI DR1, with none of the 10 matched sources exhibiting a redshift difference greater than 0.01, except for J1216+7948 at $z=1.68$. For this object, the redshift difference is $\Delta z = 0.016$, which can be attributed to the Mg\,\textsc{ii} emission line being affected by telluric absorption, making its peak difficult to identify precisely.

Most of our quasars (48 out of 62) are brighter than the DESI lower magnitude cut of $r = 16.5$~AB mag. However, all 10 overlapping quasars reported in the DESI catalog do not satisfy the $r > 16.5$~AB mag criterion based on our spectroscopic flux levels. Given the intrinsic variability of quasars, photometric fluxes may vary across observing epochs, making it difficult to determine why certain objects were excluded from the DESI candidate list.

\section{Summary and Conclusion} \label{sec:conclu}

We present the second installment of the All-sky BRIght, Complete Quasar Survey (AllBRICQS), targeting the identification of optically bright quasars in the Northern Hemisphere. Utilizing all-sky data from \textit{WISE} and \textit{Gaia} DR3, we applied the same four efficient selection criteria from \citet{onken_allbricqs_2023} and conducted spectroscopic observations of 75 candidates using the BOAO 1.8m, LJO 2.4m, and XLO 2.16m telescopes. These observations resulted in the confirmation of 62 quasars, extending the success of the AllBRICQS methodology to the northern sky.

The newly discovered quasars span redshifts of $0.09 < z < 2.48$ and bolometric luminosities from $\log L_{\mathrm{bol}} = 44.9$ to $48.0$~erg~s$^{-1}$, placing them at the luminous end of the quasar population. The purity of our Priority 1 sample, defined by $W1-W2 > 0.3$~mag, is exceptionally high at 98.4\%. Most contaminants are associated with the Priority 2 sample, which targets sources deeper within the stellar locus. As shown in Figure~\ref{fig:PrlxErr_pm}, we demonstrate that a more stringent cut on proper motion ($\mu < 0.2$~mas~yr$^{-1}$) and parallax-to-error ratio ($\varpi / \sigma_{\varpi} < 1.75$) can further improve selection purity.

Our analysis also highlights the limitations of current catalogs such as Quaia and SDSS DR16Q in recovering bright quasars. Quaia fails to include 12 of our 62 confirmed quasars and shows significant redshift discrepancies in several cases. Similarly, 18 of our new quasars are absent from SDSS DR16Q despite lying within its imaging footprint and satisfying its brightness criteria. These omissions emphasize the continued need for spectroscopic confirmation to achieve a truly complete census of luminous quasars.

Among our discoveries are rare and notable objects, including the most luminous FeLoBAL quasar to date. The AllBRICQS-North sample will serve as a valuable resource for studies of quasar physics, black hole growth, feedback processes, and host galaxy properties, particularly in regimes inaccessible to deeper but narrower surveys.

In summary, this work affirms the effectiveness of the AllBRICQS selection method and reinforces the importance of spectroscopic follow-up in uncovering the full population of the Universe’s brightest active galactic nuclei.

\section*{Acknowledgments} \label{sec:acknow}

This work was supported by the National Research Foundation of Korea (NRF) grant No. 2021M3F7A1084525, funded by the Korean government (MSIT). We also made use of data obtained at the Bohyunsan Optical Astronomy Observatory, which is operated by the Korea Astronomy and Space Science Institute.

We acknowledge the support of the staff of the Xinglong 2.16m telescope. This work was partially supported by the National Astronomical Observatories, Chinese Academy of Sciences. We also acknowledge the support of the staff of the Lijiang 2.4m telescope. Funding for the Lijiang telescope has been provided by the Chinese Academy of Sciences and the People’s Government of Yunnan Province.

This work has made use of data from the European Space Agency (ESA) mission \textit{Gaia} (\url{https://www.cosmos.esa.int/gaia}), processed by the \textit{Gaia} Data Processing and Analysis Consortium (DPAC, \url{https://www.cosmos.esa.int/web/gaia/dpac/consortium}). Funding for DPAC has been provided by national institutions, in particular the institutions participating in the \textit{Gaia} Multilateral Agreement.

This publication also makes use of data products from the Wide-field Infrared Survey Explorer (WISE), which is a joint project of the University of California, Los Angeles, and the Jet Propulsion Laboratory/California Institute of Technology, funded by the National Aeronautics and Space Administration (NASA).

We additionally used data obtained with the Samuel Oschin 48-inch and the 60-inch telescopes at Palomar Observatory as part of the Zwicky Transient Facility (ZTF) project. ZTF is supported by the National Science Foundation under Grant No. AST-1440341 and by a collaboration including Caltech, IPAC, the Weizmann Institute for Science, the Oskar Klein Center at Stockholm University, the University of Maryland, the University of Washington (UW), Deutsches Elektronen-Synchrotron and Humboldt University, Los Alamos National Laboratories, the TANGO Consortium of Taiwan, the University of Wisconsin at Milwaukee, and Lawrence Berkeley National Laboratory. Operations are conducted by Caltech Optical Observatories, IPAC, and UW.

Y.C., M.I., S.W.C., H.C., M.J., and J.H.K. acknowledge support from the NRF grant No. 2021M3F7A1084525.  
S.W.C. also acknowledges support from the Basic Science Research Program through the NRF, funded by the Ministry of Education (grant No. RS-2023-00245013).  
H.C. acknowledges support from NRF grant No. RS-2025-00573214 funded by the Korean government (MSIT).
J.H.K. also acknowledges support from NRF grant No. 2020R1A2C3011091 and from the Institute of Information \& Communications Technology Planning \& Evaluation (IITP) grant No. RS-2021-II212068, funded by the Korean government (MSIT).  
G.L. acknowledges support from the Basic Science Research Program through the NRF, funded by MSIT (grant No. 2022R1A6A3A01085930).  
Y.K. was supported by NRF grant No. 2021R1C1C2091550, funded by the Korean government (MSIT).  
G.S.H.P. acknowledges support from the Pan-STARRS project, which is operated by the Institute for Astronomy of the University of Hawaii and supported by NASA’s Near-Earth Object Observation Program under grants 80NSSC18K0971, NNX14AM74G, NNX12AR65G, NNX13AQ47G, NNX08AR22G, and 80NSSC21K1572, as well as by the State of Hawaii.  
T.K. and J.S. are supported by the NRF grant funded by the Korean government (MSIT) (grant No. RS-2023-00210597).  
D.K. acknowledges support from NRF grant No. 2021R1C1C1013580, funded by the Korean government (MSIT).

\section*{Data Availability} \label{sec:github}

The spectral data and tables presented in this paper are publicly available. An example of the spectrum data is provided in Appendix~\ref{sec:example_spec}.

\bibliography{00_2024_AllBRICQS_bib}{}
\bibliographystyle{aasjournal}

\clearpage
\newcolumntype{C}[1]{>{\centering\arraybackslash}p{#1}} 

\begin{longtable}{
p{1.7cm} C{1.0cm} C{1.0cm} C{0.7cm} C{0.7cm} C{0.7cm} C{1.7cm} C{2.3cm} C{0.7cm} C{1.2cm} C{1cm}
}

\caption{AllBRICQS Confirmed Quasars \label{tab:AllCofQuasar}} \\
\hline
\multicolumn{1}{c}{Target} &
\multicolumn{1}{c}{RA (J2000)} &
\multicolumn{1}{c}{Dec (J2000)} &
\multicolumn{1}{c}{$B_P$} &
\multicolumn{1}{c}{$R_P$} &
\multicolumn{1}{c}{$W12$} &
\multicolumn{1}{c}{$\mu$} &
\multicolumn{4}{c}{Redshift} \\
\cline{8-11}
& (deg) & (deg) & (mag) & (mag) & (mag) & (mas yr$^{-1}$)
& AllBRICQS & Quaia & CatNorth & Others \\
\hline
\endfirsthead

\hline
\multicolumn{11}{c}{\textit{Continued from previous page}} \\
\hline
\multicolumn{1}{c}{Target} &
\multicolumn{1}{c}{RA (J2000)} &
\multicolumn{1}{c}{Dec (J2000)} &
\multicolumn{1}{c}{$B_P$} &
\multicolumn{1}{c}{$R_P$} &
\multicolumn{1}{c}{$W12$} &
\multicolumn{1}{c}{$\mu$} &
\multicolumn{4}{c}{Redshift} \\
\cline{8-11}
& (deg) & (deg) & (mag) & (mag) & (mag) & (mas yr$^{-1}$)
& AllBRICQS & Quaia & CatNorth & Others \\
\hline
\endhead

\hline
\multicolumn{11}{c}{\textit{Continued on next page}} \\
\hline
\endfoot

\hline
\endlastfoot

J0002+2529 & 0.6584 & 25.4906 & 16.22 & 15.20 & 1.14 & $0.04 \pm 0.05$ & $0.3637 \pm 0.0002$ & 0.365 & 0.435 & $0.37^{1}$ \\
J0052+4721* & 13.1842 & 47.3663 & 16.89 & 15.98 & 1.11 & $0.03 \pm 0.07$ & $1.5267 \pm 0.0006$ & - & - & -\\
J0105+5137 & 16.4816 & 51.6293 & 16.63 & 15.86 & 1.25 & $0.06 \pm 0.06$ & $0.7692 \pm 0.0002$ & 0.539 & 0.790 & - \\
J0139+4036 & 24.7536 & 40.6120 & 16.23 & 15.57 & 0.73 & $0.02 \pm 0.07$ & $2.4860 \pm 0.0008$ & 1.154 & 2.209 & - \\
J0141+7307 & 25.3294 & 73.1285 & 17.29 & 15.83 & 1.19 & $0.04 \pm 0.06$ & $1.5869 \pm 0.0007$ & - & 1.698 & - \\
J0152+3143 & 28.0699 & 31.7171 & 16.59 & 16.00 & 1.47 & $0.19 \pm 0.09$ & $1.2164 \pm 0.0006$ & - & - & $1.20^{3}$ \\
J0212+3306 & 33.2025 & 33.1120 & 16.63 & 15.95 & 1.04 & $0.11 \pm 0.10$ & $0.1874 \pm 0.0001$ & 0.199 & 0.186 & - \\
J0214+3824 & 33.6604 & 38.4148 & 16.55 & 15.95 & 1.39 & $0.13 \pm 0.10$ & $1.5698 \pm 0.0011$ & 1.567 & 1.557 & $1.57^{1}$ \\
J0245+7718 & 41.3216 & 77.3059 & 17.25 & 15.92 & 0.99 & $0.15 \pm 0.08$ & $1.9793 \pm 0.0011$ & - & 2.027 & - \\
J0303+2629 & 45.7550 & 26.4884 & 16.80 & 15.95 & 1.25 & $0.04 \pm 0.10$ & $0.9068 \pm 0.0001$ & 0.654 & 0.916 & $0.91^{1}$ \\
J0324+2432 & 51.1758 & 24.5459 & 16.20 & 15.30 & 1.04 & $0.09 \pm 0.08$ & $0.2472 \pm 0.0001$ & 0.245 & 0.473 & - \\
J0503+8351 & 75.7822 & 83.8612 & 16.56 & 15.72 & 1.04 & $0.12 \pm 0.06$ & $0.2590 \pm 0.0001$ & 0.274 & 0.299 & - \\
J0528+6633 & 82.1957 & 66.5598 & 16.57 & 15.81 & 0.95 & $0.03 \pm 0.04$ & $0.6793 \pm 0.0002$ & 0.647 & 1.076 & - \\
J0602+4743 & 90.5840 & 47.7324 & 16.70 & 15.87 & 1.07 & $0.06 \pm 0.07$ & $0.2316 \pm 0.0001$ & 0.230 & 0.307 & - \\
J0633+6225 & 98.2560 & 62.4288 & 16.27 & 15.57 & 1.07 & $0.02 \pm 0.05$ & $0.3343 \pm 0.0001$ & 0.335 & 0.341 & - \\
J0643+5044 & 100.9192 & 50.7418 & 15.82 & 15.01 & 1.03 & $0.09 \pm 0.05$ & $0.2093 \pm 0.0001$ & - & 0.230 & - \\
J0653+6716 & 103.3123 & 67.2795 & 16.14 & 15.37 & 1.00 & $0.03 \pm 0.05$ & $0.2925 \pm 0.0002$ & 0.293 & 0.318 & - \\
J0707+2821 & 106.9702 & 28.3518 & 16.04 & 15.40 & 1.01 & $0.09 \pm 0.07$ & $1.9091 \pm 0.0007$ & 1.891 & 1.885 & - \\
J0708+7555 & 107.0917 & 75.9217 & 16.38 & 15.76 & 1.30 & $0.02 \pm 0.05$ & $1.8851 \pm 0.0012$ & 1.883 & 1.864 & - \\
J0708+7822 & 107.0212 & 78.3713 & 15.85 & 15.21 & 0.85 & $0.06 \pm 0.04$ & $0.2512 \pm 0.0001$ & 0.400 & 0.501 & - \\
J0728+2056 & 112.1122 & 20.9385 & 16.86 & 15.88 & 0.91 & $0.08 \pm 0.08$ & $1.3473 \pm 0.0017$ & - & 2.569 & - \\
J0750+7751 & 117.7125 & 77.8504 & 16.11 & 15.70 & 1.17 & $0.05 \pm 0.05$ & $0.7693 \pm 0.0003$ & 0.703 & 0.751 & $0.77^{1}$ \\
J0807+7550 & 121.8778 & 75.8388 & 15.66 & 14.82 & 1.18 & $0.02 \pm 0.03$ & $0.5753 \pm 0.0002$ & 0.402 & 0.847 & $0.57^{1}$ \\
J0827+5703 & 126.9324 & 57.0606 & 17.20 & 15.85 & 1.16 & $0.08 \pm 0.07$ & $1.55 \pm 0.01$ & - & 1.690 & - \\
J0919+3557 & 139.9281 & 35.9514 & 16.78 & 15.59 & 0.98 & $0.03 \pm 0.08$ & $2.365 \pm 0.001$ & - & - & $3.30^{3}$ \\
J0919+7539 & 139.9001 & 75.6623 & 16.38 & 15.89 & 1.16 & $0.06 \pm 0.06$ & $0.5238 \pm 0.0002$ & 0.517 & 0.537 & $0.52^{1}$ \\
J0938+4927* & 144.7180 & 49.4584 & 16.38 & 15.91 & 1.20 & $0.12 \pm 0.06$ & $0.7370 \pm 0.0002$ & 0.514 & - & - \\
J1143+7727 & 175.8681 & 77.4592 & 16.56 & 16.00 & 1.33 & $0.13 \pm 0.08$ & $0.9969 \pm 0.0001$ & 0.966 & 1.068 & $\text{--}^{2}$ \\
J1203+7226 & 180.7783 & 72.4370 & 16.15 & 15.31 & 0.95 & $0.10 \pm 0.06$ & $0.1033 \pm 0.0001$ & 0.226 & 0.177 & - \\
J1216+7948 & 184.1664 & 79.8003 & 16.73 & 15.94 & 1.45 & $0.00 \pm 0.07$ & $1.6834 \pm 0.0011$ & 1.697 & 1.685 & - \\
J1256+5404 & 194.1117 & 54.0742 & 16.54 & 16.00 & 0.97 & $0.05 \pm 0.06$ & $0.1993 \pm 0.0001$ & 0.279 & 0.280 & - \\
J1317+8637 & 199.3724 & 86.6268 & 16.95 & 15.92 & 1.06 & $0.08 \pm 0.06$ & $0.2414 \pm 0.0001$ & 0.240 & 0.530 & - \\
J1336+7550 & 204.0408 & 75.8430 & 15.48 & 14.93 & 1.02 & $0.04 \pm 0.04$ & $0.2351 \pm 0.0001$ & 0.400 & 0.237 & $0.23^{1}$ \\
J1356+3840 & 209.0514 & 38.6727 & 17.98 & 15.60 & 1.11 & $0.08 \pm 0.05$ & $1.39 \pm 0.01$ & - & 2.679 & - \\
J1410+7533 & 212.5899 & 75.5651 & 16.43 & 15.96 & 1.29 & $0.02 \pm 0.07$ & $0.852 \pm 0.001$ & 2.701 & 0.960 & $3.26^{1}$ \\
J1505+8604 & 226.3111 & 86.0753 & 16.53 & 15.76 & 1.40 & $0.05 \pm 0.06$ & $1.2333 \pm 0.0008$ & 1.219 & 1.219 & - \\
J1600+6651 & 240.2484 & 66.8511 & 16.47 & 16.25 & 1.18 & $0.09 \pm 0.08$ & $0.5137 \pm 0.0002$ & 0.578 & 0.604 & - \\
J1649+4726 & 252.4820 & 47.4452 & 15.57 & 15.05 & 1.06 & $0.04 \pm 0.04$ & $0.2726 \pm 0.0001$ & 0.400 & 0.297 & - \\
J1728+4303 & 262.0179 & 43.0590 & 16.30 & 15.70 & 1.39 & $0.02 \pm 0.05$ & $1.5429 \pm 0.0006$ & 1.548 & 1.547 & $1.60^{3}$ \\
J1741+3026 & 265.4283 & 30.4466 & 16.70 & 15.97 & 1.08 & $0.03 \pm 0.06$ & $0.2039 \pm 0.0001$ & 0.202 & 0.178 & $0.20^{1}$ \\
J1745+5414 & 266.3538 & 54.2412 & 16.55 & 15.70 & 1.28 & $0.05 \pm 0.06$ & $1.3567 \pm 0.0004$ & 1.356 & 1.348 & $1.40^{3}$ \\
J1807+3637 & 271.7788 & 36.6215 & 16.44 & 15.90 & 1.19 & $0.07 \pm 0.06$ & $1.1141 \pm 0.0006$ & 1.104 & 1.092 & - \\
J1809+2758 & 272.4902 & 27.9709 & 17.50 & 15.97 & 0.76 & $0.04 \pm 0.06$ & $2.41 \pm 0.01$ & - & 3.135 & - \\
J1811+2409 & 272.8511 & 24.1503 & 16.63 & 15.97 & 1.10 & $0.06 \pm 0.05$ & $0.9215 \pm 0.0002$ & 0.904 & 0.899 & - \\
J1813+3844 & 273.4528 & 38.7482 & 15.95 & 15.42 & 1.03 & $0.09 \pm 0.05$ & $0.2432 \pm 0.0001$ & 0.318 & 0.312 & $0.24^{1}$ \\
J1813+4531* & 273.3915 & 45.5216 & 16.60 & 15.80 & 1.11 & $0.05 \pm 0.07$ & $0.0931 \pm 0.0001$ & 0.238 & - & - \\
J1846+3616 & 281.7226 & 36.2796 & 16.69 & 15.95 & 0.98 & $0.03 \pm 0.06$ & $0.1340 \pm 0.0001$ & 0.197 & 0.152 & - \\
J1846+8425 & 281.7071 & 84.4181 & 15.72 & 15.08 & 0.93 & $0.05 \pm 0.04$ & $0.2255 \pm 0.0001$ & 0.458 & 0.416 & $0.22^{1}$ \\
J1859+5130 & 284.9838 & 51.5094 & 16.49 & 15.94 & 1.27 & $0.07 \pm 0.06$ & $0.7117 \pm 0.0004$ & 0.710 & 0.739 & - \\
J1918+4222 & 289.5642 & 42.3799 & 16.38 & 15.86 & 1.06 & $0.04 \pm 0.05$ & $0.4397 \pm 0.0001$ & 0.485 & 0.502 & - \\
J1938+5408 & 294.5447 & 54.1489 & 16.57 & 15.79 & 1.08 & $0.06 \pm 0.06$ & $0.2717 \pm 0.0001$ & 0.273 & 0.289 & $1.00^{4}$ \\
J2114+2524 & 318.5876 & 25.4062 & 16.32 & 15.59 & 1.11 & $0.04 \pm 0.07$ & $0.0897 \pm 0.0001$ & 0.239 & 0.205 & - \\
J2115+2526* & 318.9501 & 25.4455 & 16.72 & 15.87 & 1.01 & $0.03 \pm 0.08$ & $0.2302 \pm 0.0001$ & 0.229 & - & - \\
J2135+3348 & 323.9296 & 33.8053 & 16.62 & 15.98 & 1.11 & $0.09 \pm 0.05$ & $0.4296 \pm 0.0003$ & 0.419 & 0.442 & - \\
J2139+2929 & 324.8727 & 29.4961 & 16.65 & 15.92 & 1.09 & $0.05 \pm 0.05$ & $0.3803 \pm 0.0001$ & 0.381 & 0.360 & - \\
J2206+2757 & 331.7160 & 27.9662 & 16.50 & 15.90 & 1.06 & $0.00 \pm 0.07$ & $0.2512 \pm 0.0001$ & 0.279 & 0.309 & $0.90^{3}$ \\
J2217+3555 & 334.3668 & 35.9300 & 16.63 & 15.85 & 1.04 & $0.10 \pm 0.07$ & $0.1466 \pm 0.0002$ & 0.158 & 0.181 & - \\
J2225+3952 & 336.4773 & 39.8795 & 16.73 & 15.97 & 1.37 & $0.04 \pm 0.05$ & $1.75 \pm 0.01$ & 1.747 & 1.745 & - \\
J2232+4230 & 338.0374 & 42.5016 & 16.60 & 15.97 & 1.34 & $0.01 \pm 0.05$ & $1.1386 \pm 0.0004$ & - & 1.490 & - \\
J2235+4431 & 338.9504 & 44.5308 & 16.70 & 15.96 & 1.22 & $0.03 \pm 0.05$ & $0.9575 \pm 0.0002$ & - & 1.462 & - \\
J2332+2136 & 353.1988 & 21.6103 & 16.39 & 15.82 & 1.32 & $0.11 \pm 0.06$ & $1.0160 \pm 0.0005$ & 0.387 & 1.045 & $0.80^{4}$ \\
J2359+8307* & 359.9554 & 83.1212 & 16.40 & 15.71 & 0.97 & $0.11 \pm 0.07$ & $0.2861 \pm 0.0001$ & 0.285 & - & - \\
\end{longtable}

\newcolumntype{C}[1]{>{\centering\arraybackslash}p{#1}} 

\begin{longtable}{
p{1.7cm} C{1.0cm} C{1.0cm} C{0.7cm} C{0.7cm} C{0.7cm} C{1.7cm} C{2.3cm} C{0.7cm} C{1.2cm} C{1cm}
}
\caption{AllBRICQS Unclassified Sources \label{tab:AllUnknown}} \\
\hline
\multicolumn{1}{c}{Target} &
\multicolumn{1}{c}{RA (J2000)} &
\multicolumn{1}{c}{Dec (J2000)} &
\multicolumn{1}{c}{$B_P$} &
\multicolumn{1}{c}{$R_P$} &
\multicolumn{1}{c}{$W12$} &
\multicolumn{1}{c}{$\mu$} &
\multicolumn{4}{c}{Redshift} \\
\cline{8-11}
& (deg) & (deg) & (mag) & (mag) & (mag) & (mas yr$^{-1}$)
& AllBRICQS & Quaia & CatNorth & Others \\
\hline
\endfirsthead

\hline
\multicolumn{11}{c}{\textit{Continued from previous page}} \\
\hline
\multicolumn{1}{c}{Target} &
\multicolumn{1}{c}{RA (J2000)} &
\multicolumn{1}{c}{Dec (J2000)} &
\multicolumn{1}{c}{$B_P$} &
\multicolumn{1}{c}{$R_P$} &
\multicolumn{1}{c}{$W12$} &
\multicolumn{1}{c}{$\mu$} &
\multicolumn{4}{c}{Redshift} \\
\cline{8-11}
& (deg) & (deg) & (mag) & (mag) & (mag) & (mas yr$^{-1}$)
& AllBRICQS & Quaia & CatNorth & Others \\
\hline
\endhead

\hline
\multicolumn{11}{c}{\textit{Continued on next page}} \\
\hline
\endfoot

\hline
\endlastfoot



J0219+4828 & 34.9536 & 48.4683 & 17.04 & 15.96 & 0.26 & $0.21 \pm 0.11$ & - & - & - & - \\
J0712+3142 & 108.2353 & 31.7060 & 17.02 & 15.74 & 1.12 & $0.06 \pm 0.09$ & 0.72 & - & - & - \\
J1040+6711 & 160.1677 & 67.1917 & 16.27 & 15.47 & 1.05 & $0.03 \pm 0.04$ & 0.74 & 0.474 & - & - \\
J1418+6702 & 214.6939 & 67.0424 & 16.20 & 15.14 & 1.48 & $0.03 \pm 0.05$ & 1.80 & - & - & - \\
J2330+3937 & 352.6161 & 39.6210 & 16.87 & 15.69 & 1.36 & $0.16 \pm 0.07$ & 1.90 & - & - & - \\

\end{longtable}

\newcolumntype{C}[1]{>{\centering\arraybackslash}p{#1}} 

\begin{longtable}{
p{1.7cm} C{1.0cm} C{1.0cm} C{0.7cm} C{0.7cm} C{0.7cm} C{1.7cm} C{2.3cm} C{0.7cm} C{1.2cm} C{1cm}
}
\caption{AllBRICQS Stars \label{tab:AllStar}} \\
\hline
\multicolumn{1}{c}{Target} &
\multicolumn{1}{c}{RA (J2000)} &
\multicolumn{1}{c}{Dec (J2000)} &
\multicolumn{1}{c}{$B_P$} &
\multicolumn{1}{c}{$R_P$} &
\multicolumn{1}{c}{$W12$} &
\multicolumn{1}{c}{$\mu$} &
\multicolumn{4}{c}{Redshift} \\
\cline{8-11}
& (deg) & (deg) & (mag) & (mag) & (mag) & (mas yr$^{-1}$)
& AllBRICQS & Quaia & CatNorth & Others \\
\hline
\endfirsthead

\hline
\multicolumn{11}{c}{\textit{Continued from previous page}} \\
\hline
\multicolumn{1}{c}{Target} &
\multicolumn{1}{c}{RA (J2000)} &
\multicolumn{1}{c}{Dec (J2000)} &
\multicolumn{1}{c}{$B_P$} &
\multicolumn{1}{c}{$R_P$} &
\multicolumn{1}{c}{$W12$} &
\multicolumn{1}{c}{$\mu$} &
\multicolumn{4}{c}{Redshift} \\
\cline{8-11}
& (deg) & (deg) & (mag) & (mag) & (mag) & (mas yr$^{-1}$)
& AllBRICQS & Quaia & CatNorth & Others \\
\hline
\endhead

\hline
\multicolumn{11}{c}{\textit{Continued on next page}} \\
\hline
\endfoot

\hline
\endlastfoot



J0022+3431 & 5.6257 & 34.5267 & 15.66 & 15.67 & -0.24 & $0.15 \pm 0.07$ & 0.00 & - & - & - \\
J0140+4353 & 25.0324 & 43.8890 & 16.50 & 15.81 & -0.03 & $0.16 \pm 0.07$ & 0.00 & - & - & - \\
J0218+4534 & 34.6586 & 45.5777 & 16.57 & 15.84 & 0.06 & $0.13 \pm 0.10$ & 0.00 & - & - & - \\
J0319+3309* & 49.9753 & 33.1636 & 15.80 & 14.68 & 0.32 & $1.00 \pm 0.38$ & 0.00 & - & - & - \\
J0422+2854 & 65.7033 & 28.9040 & 16.67 & 15.62 & 0.04 & $1.07 \pm 0.65$ & 0.00 & - & - & - \\
J0425+3454 & 66.2837 & 34.9133 & 17.13 & 15.91 & 0.05 & $0.25 \pm 0.10$ & 0.00 & - & - & - \\
J0536+5548 & 84.2075 & 55.8103 & 17.05 & 15.97 & -0.03 & $0.15 \pm 0.08$ & 0.00 & - & - & - \\
J2357+4826 & 359.3479 & 48.4369 & 15.67 & 14.84 & 0.00 & $0.92 \pm 0.35$ & 0.00 & - & - & - \\

\end{longtable}


\begin{flushleft}
\textbf{Notes.} \\
Table~\ref{tab:AllCofQuasar},\ref{tab:AllUnknown} and \ref{tab:AllStar} provide a comprehensive overview of the confirmed AllBRICQS quasars, unclassified sources and stars, listed in ascending order of RA. The $B_P$,  $R_P$ magnitudes and proper motion ($\mu$) are sourced from Gaia DR3 \citep{gaia_collaboration_gaia_2016, gaia_collaboration_gaia_2021}, while the W1-W2 color ($W_{12}$) is derived from CatWISE2020 \citep{marocco_catwise2020_2021} and supplemented with AllWISE data when unavailable \citep{wright_wide-field_2010}. Although the initial candidate selection was based on AllWISE photometry, we report CatWISE values here to reflect the intended selection method. As a result, some sources may appear to lie below the $W1-W2$ color cut at 0.2. Targets lacking CatWISE data are marked with an asterisk next to the target name.

In the redshift columns, the “AllBRICQS” entry indicates our spectroscopic redshift measurements. The “Quaia” column lists estimated redshifts from the Gaia-unWISE quasar candidate catalog \citep{storey-fisher_quaia_2024}, computed using a $k$-nearest neighbors (KNN) model. The “CatNorth” column presents estimated redshifts from the improved Gaia DR3 quasar candidate catalog \citep{fu_catnorth_2024}. Additional redshift values from various literature sources are also included, with numerical superscripts indicating the reference source, as follows: \\

1. Gaia DR3 QSO candidates \citep{gaia_collaboration_gaia_2023-1}, \url{https://gea.esac.esa.int/archive}\\
2. VizieR Online Data Catalog: Lick Northern Proper Motion: NPM1 Ref. Galaxies \citep{klemola_vizier_1994}\\
3. Bayesian High-redshift Quasar Classification \citep{richards_bayesian_2015} \\
4. Identification of 1.4 Million AGN using WISE Data \citep{secrest_identification_2015}
\end{flushleft}

\appendix

\section{ZTF magnitudes of AllBRICQS Quasar \label{sec:ZTFmagQuasar}}

Table~\ref{tab:Quasarztf} presents the synthetic photometry results in the ZTF DR22 $g$- and $r$-filters \citep{masci_zwicky_2019, bellm_zwicky_2019}. A detailed explanation of the flux scaling process is provided in Section~\ref{sec:flscl}. For J0324+2432, J0602+4743, J0633+6225, and J2115+2526, which were observed with XLT, the spectra were scaled using the ZTF $g$-filter magnitudes.

\newcolumntype{C}[1]{>{\centering\arraybackslash}p{#1}} 

\begin{longtable}{
p{1.7cm} C{1.5cm} C{1.5cm} C{2.0cm} C{2.0cm}
}
\caption{ZTF synthetic photometry magnitude of Quasars} \label{tab:Quasarztf} \\
\hline
Target & RA (J2000) & Dec (J2000) & $g$ & $r$ \\
 & ($\mathrm{deg}$) & ($\mathrm{deg}$) & (AB mag) & (AB mag) \\
\hline
\endfirsthead

\hline
\multicolumn{5}{c}{\textit{Continued from previous page}} \\
\hline
Target & RA (J2000) & Dec (J2000) & $g$ & $r$ \\

\hline
\endhead

\hline
\multicolumn{5}{c}{\textit{Continued on next page}} \\
\hline
\endfoot

\hline
\endlastfoot

J0002+2529 & 0.6584 & 25.4906 & $16.241 \pm 0.067$ & $15.488 \pm 0.034$ \\
J0052+4721 & 13.1842 & 47.3663 & $16.895 \pm 0.104$ & $16.423 \pm 0.064$ \\
J0105+5137 & 16.4816 & 51.6293 & $16.972 \pm 0.150$ & $16.496 \pm 0.086$ \\
J0139+4036 & 24.7536 & 40.6120 & $16.210 \pm 0.069$ & $16.278 \pm 0.060$ \\
J0141+7307 & 25.3294 & 73.1285 & $17.224 \pm 0.221$ & $16.300 \pm 0.077$ \\
J0152+3143 & 28.0699 & 31.7171 & $16.794 \pm 0.071$ & $16.366 \pm 0.039$ \\
J0212+3306 & 33.2025 & 33.1120 & $17.380 \pm 0.125$ & $17.207 \pm 0.102$ \\
J0214+3824 & 33.6604 & 38.4148 & $16.642 \pm 0.048$ & $16.450 \pm 0.040$ \\
J0245+7718 & 41.3216 & 77.3059 & $17.687 \pm 0.127$ & $16.700 \pm 0.052$ \\
J0303+2629 & 45.7550 & 26.4884 & $16.676 \pm 0.054$ & $16.333 \pm 0.033$ \\
J0324+2432 & 51.1758 & 24.5459 & $16.667 \pm 0.143$ & $16.162 \pm 0.086$ \\
J0503+8351 & 75.7822 & 83.8612 & $16.627 \pm 0.060$ & $16.266 \pm 0.036$ \\
J0528+6633 & 82.1957 & 66.5598 & $16.794 \pm 0.066$ & $16.39 \pm 0.038$ \\
J0602+4743 & 90.5840 & 47.7324 & $16.926 \pm 0.268$ & $16.500 \pm 0.162$ \\
J0633+6225 & 98.2560 & 62.4288 & $16.602 \pm 0.146$ & $16.285 \pm 0.101$ \\
J0643+5044 & 100.9192 & 50.7418 & $15.942 \pm 0.037$ & $15.661 \pm 0.025$ \\
J0653+6716 & 103.3123 & 67.2795 & $16.264 \pm 0.047$ & $15.969 \pm 0.030$ \\
J0707+2821 & 106.9702 & 28.3518 & $16.290 \pm 0.113$ & $15.936 \pm 0.075$ \\
J0708+7555 & 107.0917 & 75.9217 & $16.575 \pm 0.057$ & $16.425 \pm 0.040$ \\
J0708+7822 & 107.0212 & 78.3713 & $15.741 \pm 0.034$ & $15.672 \pm 0.026$ \\
J0728+2056 & 112.1122 & 20.9385 & $17.162 \pm 0.211$ & $16.428 \pm 0.100$ \\
J0750+7751 & 117.7125 & 77.8504 & $16.148 \pm 0.043$ & $16.206 \pm 0.035$ \\
J0807+7550 & 121.8778 & 75.8388 & $15.904 \pm 0.037$ & $15.593 \pm 0.024$ \\
J0827+5703 & 126.9324 & 57.0606 & $17.195 \pm 0.128$ & $16.548 \pm 0.062$ \\
J0919+3557 & 139.9281 & 35.9514 & $17.013 \pm 0.076$ & $16.314 \pm 0.038$ \\
J0919+7539 & 139.9001 & 75.6623 & $16.286 \pm 0.043$ & $16.211 \pm 0.035$ \\
J0938+4927 & 144.7180 & 49.4584 & $16.364 \pm 0.045$ & $16.329 \pm 0.038$ \\
J1143+7727 & 175.8681 & 77.4592 & $16.752 \pm 0.059$ & $16.449 \pm 0.041$ \\
J1203+7226 & 180.7783 & 72.4370 & $16.141 \pm 0.040$ & $15.895 \pm 0.030$ \\
J1216+7948 & 184.1664 & 79.8003 & $17.373 \pm 0.093$ & $16.697 \pm 0.047$ \\
J1256+5404 & 194.1117 & 54.0742 & $16.596 \pm 0.055$ & $16.364 \pm 0.040$ \\
J1317+8637 & 199.3724 & 86.6268 & $16.904 \pm 0.072$ & $16.346 \pm 0.043$ \\
J1336+7550 & 204.0408 & 75.8430 & $15.641 \pm 0.027$ & $15.498 \pm 0.021$ \\
J1356+3840 & 209.0514 & 38.6727 & $18.653 \pm 0.273$ & $16.697 \pm 0.047$ \\
J1410+7533 & 212.5899 & 75.5651 & $16.332 \pm 0.040$ & $16.438 \pm 0.035$ \\
J1505+8604 & 226.3111 & 86.0753 & $16.331 \pm 0.047$ & $16.030 \pm 0.035$ \\
J1600+6651 & 240.2484 & 66.8511 & $16.561 \pm 0.050$ & $16.499 \pm 0.040$ \\
J1649+4726 & 252.4820 & 47.4452 & $15.784 \pm 0.031$ & $15.600 \pm 0.023$ \\
J1728+4303 & 262.0179 & 43.0590 & $16.688 \pm 0.054$ & $16.259 \pm 0.034$ \\
J1741+3026 & 265.4283 & 30.4466 & $16.795 \pm 0.059$ & $16.531 \pm 0.040$ \\
J1745+5414 & 266.3538 & 54.2412 & $17.263 \pm 0.082$ & $16.271 \pm 0.034$ \\
J1807+3637 & 271.7788 & 36.6215 & $17.138 \pm 0.075$ & $16.577 \pm 0.041$ \\
J1809+2758 & 272.4902 & 27.9709 & $18.154 \pm 0.035$ & $16.840 \pm 0.006$ \\
J1811+2409 & 272.8511 & 24.1503 & $16.753 \pm 1.236$ & $16.300 \pm 0.767$ \\
J1813+3844 & 273.4528 & 38.7482 & $16.117 \pm 0.068$ & $16.062 \pm 0.061$ \\
J1813+4531 & 273.3915 & 45.5216 & $17.040 \pm 0.072$ & $16.160 \pm 0.033$ \\
J1846+3616 & 281.7226 & 36.2796 & $17.484 \pm 0.188$ & $16.753 \pm 0.104$ \\
J1846+8425 & 281.7071 & 84.4181 & $15.931 \pm 0.037$ & $15.713 \pm 0.029$ \\
J1859+5130 & 284.9838 & 51.5094 & $16.306 \pm 0.077$ & $16.187 \pm 0.065$ \\
J1918+4222 & 289.5642 & 42.3799 & $16.265 \pm 1.085$ & $16.200 \pm 0.960$ \\
J1938+5408 & 294.5447 & 54.1489 & $16.570 \pm 0.073$ & $16.383 \pm 0.059$ \\
J2114+2524 & 318.5876 & 25.4062 & $16.378 \pm 0.068$ & $16.038 \pm 0.049$ \\
J2115+2526 & 318.9501 & 25.4455 & $17.142 \pm 0.276$ & $16.632 \pm 0.174$ \\
J2135+3348 & 323.9296 & 33.8053 & $16.363 \pm 0.046$ & $16.260 \pm 0.041$ \\
J2139+2929 & 324.8727 & 29.4961 & $16.611 \pm 0.054$ & $16.369 \pm 0.044$ \\
J2206+2757 & 331.7160 & 27.9662 & $16.940 \pm 0.031$ & $16.700 \pm 0.025$ \\
J2217+3555 & 334.3668 & 35.9300 & $16.651 \pm 0.079$ & $16.467 \pm 0.065$ \\
J2225+3952 & 336.4773 & 39.8795 & $16.873 \pm 0.058$ & $16.600 \pm 0.045$ \\
J2232+4230 & 338.0374 & 42.5016 & $16.920 \pm 0.068$ & $16.550 \pm 0.038$ \\
J2235+4431 & 338.9504 & 44.5308 & $16.854 \pm 0.085$ & $16.450 \pm 0.060$ \\
J2332+2136 & 353.1988 & 21.6103 & $16.578 \pm 0.092$ & $16.328 \pm 0.071$ \\
J2359+8307 & 359.9554 & 83.1212 & $16.674 \pm 0.060$ & $16.453 \pm 0.040$ \\
\end{longtable}

\clearpage
\section{AllBRICQS Luminosities \label{sec:LumQuasar}}

In Table~\ref{tab:QuasarBLum}, we present the continuum and bolometric luminosities for the confirmed AllBRICQS quasars, arranged in ascending order of right ascension (RA). Two quasars, J0633+6225 and J1809+2758, are not included in the table due to insufficient wavelength coverage for reliable luminosity estimation. Only valid data points are used in the analysis, and when multiple continuum luminosities are available, we adopt the mean value. Additional details on the bolometric luminosity calculation are provided in Section~\ref{sec:bollum}.

For J1859+5130 at $z = 0.71$, although both the 3000\,\AA\ and 5100\,\AA\ continuum regions are covered, only the 3000\,\AA\ luminosity is used in the bolometric correction. This decision is based on unreliable flux measurements at $\lambda \gtrsim 7500$\,\AA\ in the spectrum obtained on UT 2022 October 27.

\newcolumntype{C}[1]{>{\centering\arraybackslash}p{#1}} 

\begin{longtable}{
p{1.7cm} C{1.5cm} C{1.5cm} C{1.5cm} C{1.5cm}
}
\caption{Bolometric and Continuum Luminosities of Quasars} \label{tab:QuasarBLum} \\
\hline
Target & $\log( L_{\mathrm{bol}})$ & $\log( L_{1450\,\text{\AA}})$ & $\log( L_{3000\,\text{\AA}})$ & $\log( L_{5100\,\text{\AA}})$ \\
 &  (erg s$^{-1}$) & (erg s$^{-1}$) & (erg s$^{-1}$) & (erg s$^{-1}$) \\
\hline
\endfirsthead

\hline
\multicolumn{5}{c}{\textit{Continued from previous page}} \\
\hline
Target & $\log( L_{\mathrm{bol}})$ & $\log( L_{1450\,\text{\AA}})$ & $\log( L_{3000\,\text{\AA}})$ & $\log( L_{5100\,\text{\AA}})$ \\
\hline
\endhead

\hline
\multicolumn{5}{c}{\textit{Continued on next page}} \\
\hline
\endfoot

\hline
\endlastfoot
J0002+2529 & 46.49 & - & - & 45.62 \\
J0052+4721 & 47.04 & - & 46.35 & - \\
J0105+5137 & 46.86 & - & 46.17 & - \\
J0139+4036 & 47.46 & 46.94 & - & - \\
J0141+7307 & 47.69 & - & 47.01 & - \\
J0152+3143 & 47.27 & - & 46.59 & - \\
J0212+3306 & 45.37 & - & - & 44.39 \\
J0214+3824 & 47.12 & - & 46.42 & - \\
J0245+7718 & 47.48 & - & 46.80 & - \\
J0303+2629 & 46.96 & - & 46.26 & - \\
J0324+2432 & 45.96 & - & - & 45.03 \\
J0503+8351 & 45.95 & - & - & 45.02 \\
J0528+6633 & 46.65 & - & - & 45.79 \\
J0602+4743 & 45.91 & - & - & 44.98 \\
J0643+5044 & 46.00 & - & - & 45.08 \\
J0653+6716 & 46.15 & - & - & 45.24 \\
J0707+2821 & 47.52 & - & 46.84 & - \\
J0708+7555 & 47.24 & - & 46.55 & - \\
J0708+7822 & 46.10 & - & - & 45.19 \\
J0728+2056 & 47.09 & - & 46.40 & - \\
J0750+7751 & 46.78 & - & 46.08 & - \\
J0807+7550 & 46.76 & - & - & 45.91 \\
J0827+5703 & 47.21 & - & 46.52 & - \\
J0919+3557 & 47.68 & 47.18 & - & - \\
J0919+7539 & 46.39 & - & - & 45.50 \\
J0938+4927 & 46.63 & - & 45.98 & 45.72 \\
J1143+7727 & 46.84 & - & 46.14 & - \\
J1203+7226 & 45.35 & - & - & 44.36 \\
J1216+7948 & 47.17 & - & 46.48 & - \\
J1256+5404 & 45.67 & - & - & 44.71 \\
J1317+8637 & 45.98 & - & - & 45.06 \\
J1336+7550 & 46.11 & - & - & 45.20 \\
J1356+3840 & 47.18 & - & 46.49 & - \\
J1410+7533 & 46.78 & - & 46.07 & - \\
J1505+8604 & 47.27 & - & 46.58 & - \\
J1600+6651 & 46.41 & - & - & 45.52 \\
J1649+4726 & 46.18 & - & - & 45.28 \\
J1728+4303 & 47.19 & - & 46.50 & - \\
J1741+3026 & 45.63 & - & - & 44.67 \\
J1745+5414 & 47.15 & - & 46.46 & - \\
J1807+3637 & 46.81 & - & 46.11 & - \\
J1811+2409 & 46.90 & - & 46.20 & - \\
J1813+3844 & 45.92 & - & - & 44.99 \\
J1813+4531 & 44.94 & - & - & 43.92 \\
J1846+3616 & 45.22 & - & - & 44.22 \\
J1846+8425 & 46.04 & - & - & 45.12 \\
J1859+5130 & 46.73 & - & 46.03 & 45.40 \\
J1918+4222 & 46.45 & - & - & 45.57 \\
J1938+5408 & 45.97 & - & - & 45.05 \\
J2114+2524 & 45.18 & - & - & 44.18 \\
J2115+2526 & 45.69 & - & - & 44.74 \\
J2135+3348 & 46.39 & - & - & 45.50 \\
J2139+2929 & 46.24 & - & - & 45.33 \\
J2206+2757 & 45.78 & - & - & 44.84 \\
J2217+3555 & 45.47 & - & - & 44.50 \\
J2225+3952 & 47.14 & - & 46.45 & - \\
J2232+4230 & 46.94 & - & 46.24 & - \\
J2235+4431 & 46.96 & - & 46.26 & - \\
J2332+2136 & 46.93 & - & 46.24 & - \\
J2359+8307 & 46.04 & - & - & 45.11 \\
\end{longtable}

\section{Gallery of AllBRICQS Observed Spectra} \label{sec:ObsSpec}
\subsection{Gallery of AllBRICQS Quasar Spectra} \label{sec:GalQuasar}

Figure~\ref{fig:QuasarSED1} presents the spectra of the AllBRICQS quasars, arranged in order of ascending redshift. Spectra with incomplete wavelength coverage correspond to targets observed with either XLT or LJT, where partial spectral data were obtained. 

\setlength{\abovecaptionskip}{-1cm}
\begin{figure*}[t]
    \centering
    \vspace{-3.5cm}  
    \includegraphics[width=\textwidth, trim=0cm 2cm 0cm 0cm, clip]{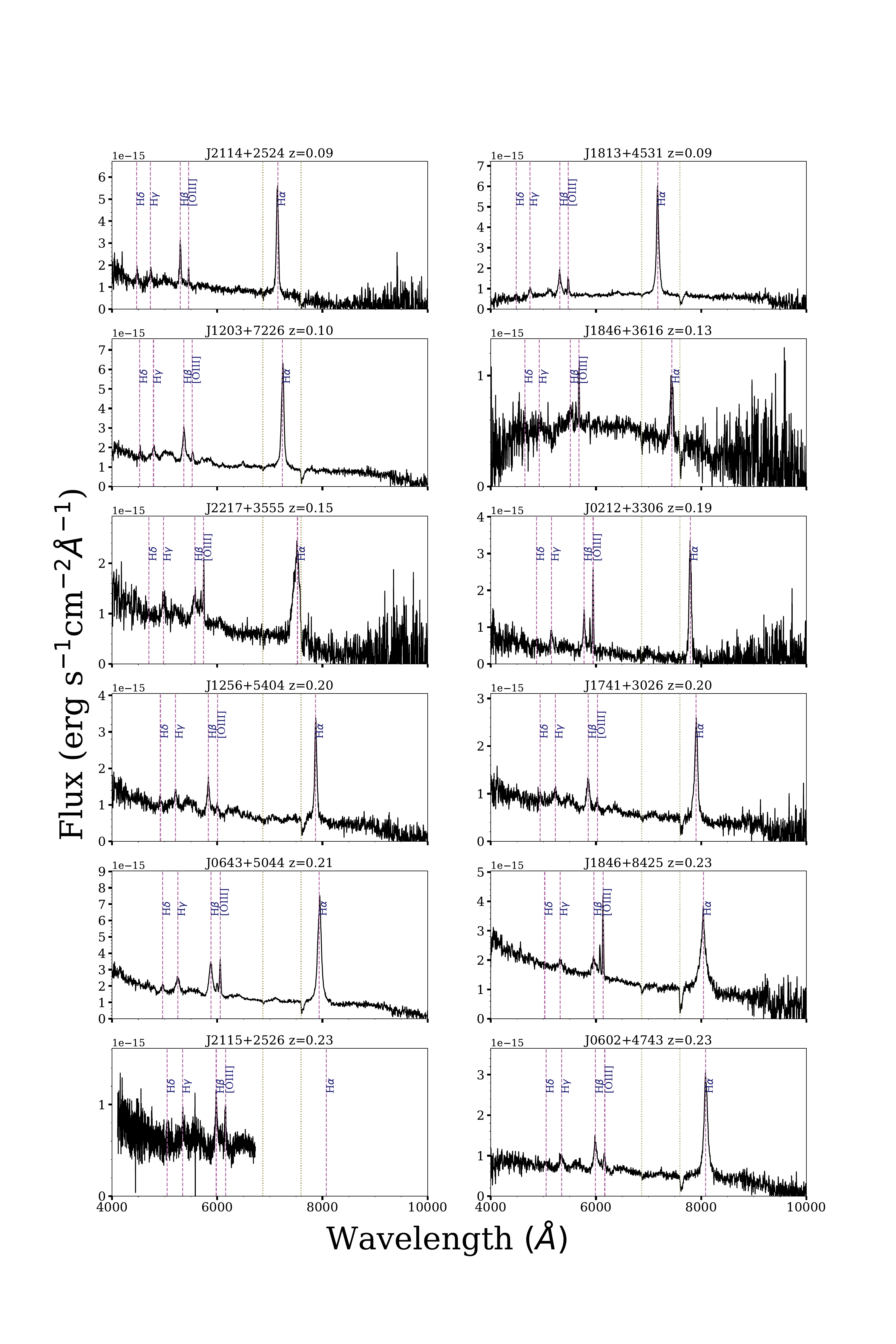}
    \caption{Spectra of the AllBRICQS Northern Hemisphere quasar sample, displayed in order of increasing redshift. Each panel is labeled with the target name and the redshift measured from the observed spectrum. The $y$-axis shows the scaled flux, aligned with contemporaneous broadband photometry. Pink dotted vertical lines indicate the positions of key quasar emission features, with line identifications shown to the right of each marker. Green dotted vertical lines at 6867.2~\AA\ and 7593.7~\AA\ mark regions affected by telluric absorption, which may impact spectral interpretation.}
    \label{fig:QuasarSED1}
\end{figure*}

\clearpage  

\begin{figure*}[t]
    \figurenum{7}
    \centering
    \vspace{-3cm}
    \includegraphics[width=\textwidth, trim=0cm 2cm 0cm 0cm, clip]{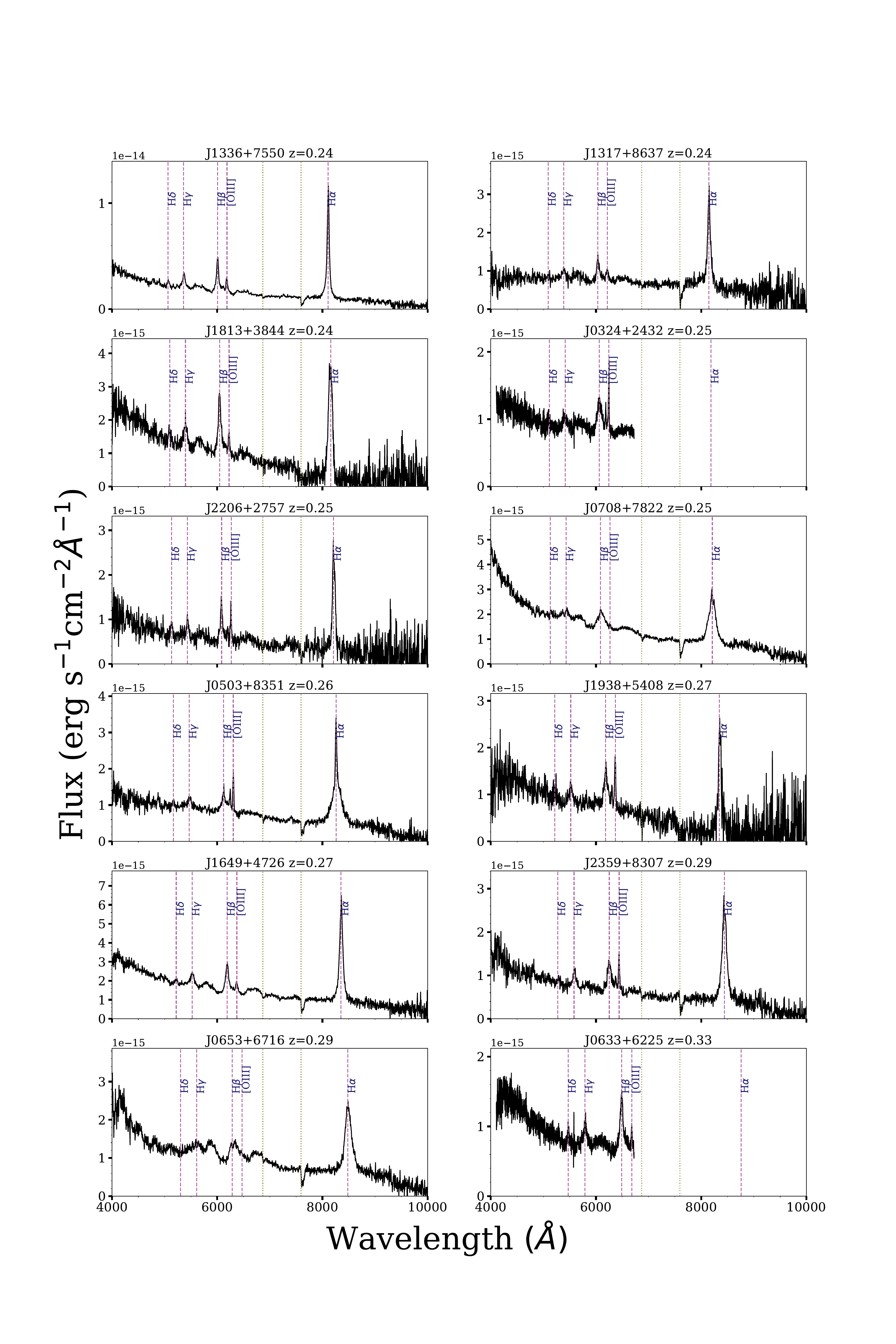}
    \caption{(Continued)}
\end{figure*}

\clearpage

\begin{figure*}[t]
    \figurenum{7}
    \centering
    \vspace{-3cm}
    \includegraphics[width=\textwidth, trim=0cm 2cm 0cm 0cm, clip]{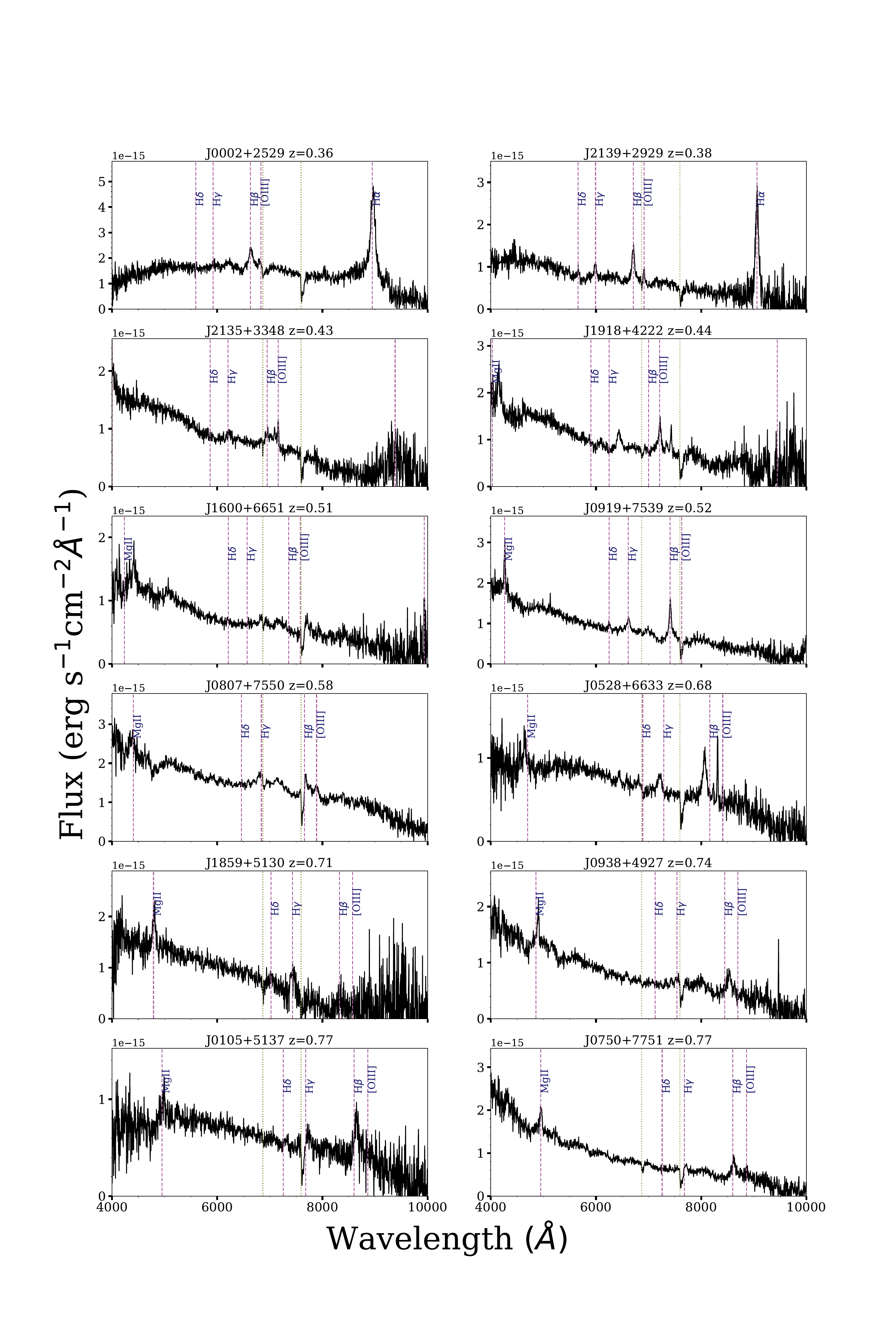}
    \caption{(Continued)}
\end{figure*}

\clearpage

\begin{figure*}[t]
    \figurenum{7}
    \centering
    \vspace{-3cm}
    \includegraphics[width=\textwidth, trim=0cm 2cm 0cm 0cm, clip]{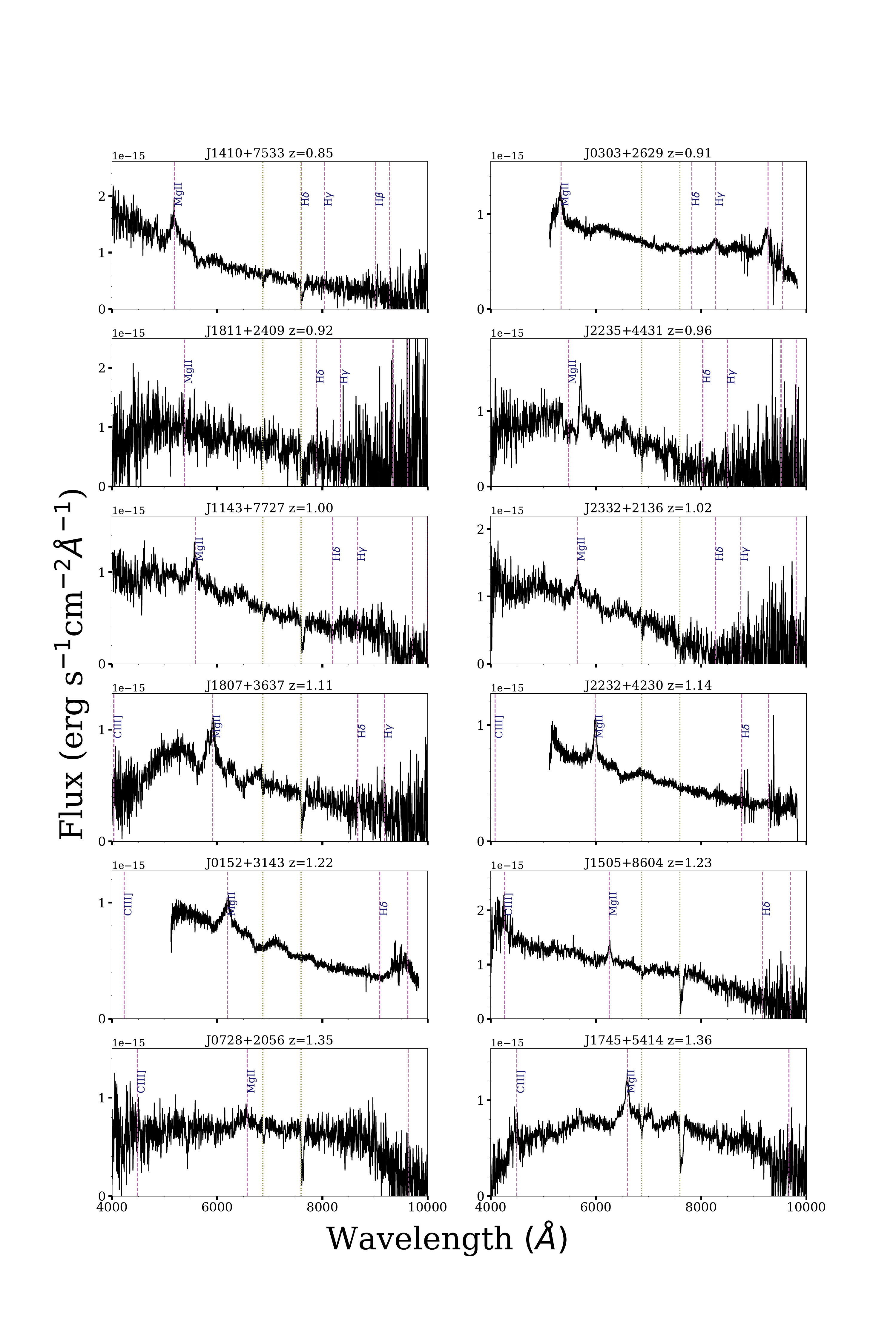}
    \caption{(Continued)}
\end{figure*}

\clearpage

\setlength{\abovecaptionskip}{-1cm}
\begin{figure*}[t]
    \figurenum{7}
    \centering
    \vspace{-5cm}
    \includegraphics[width=\textwidth, trim=0cm 2cm 0cm 0cm, clip]{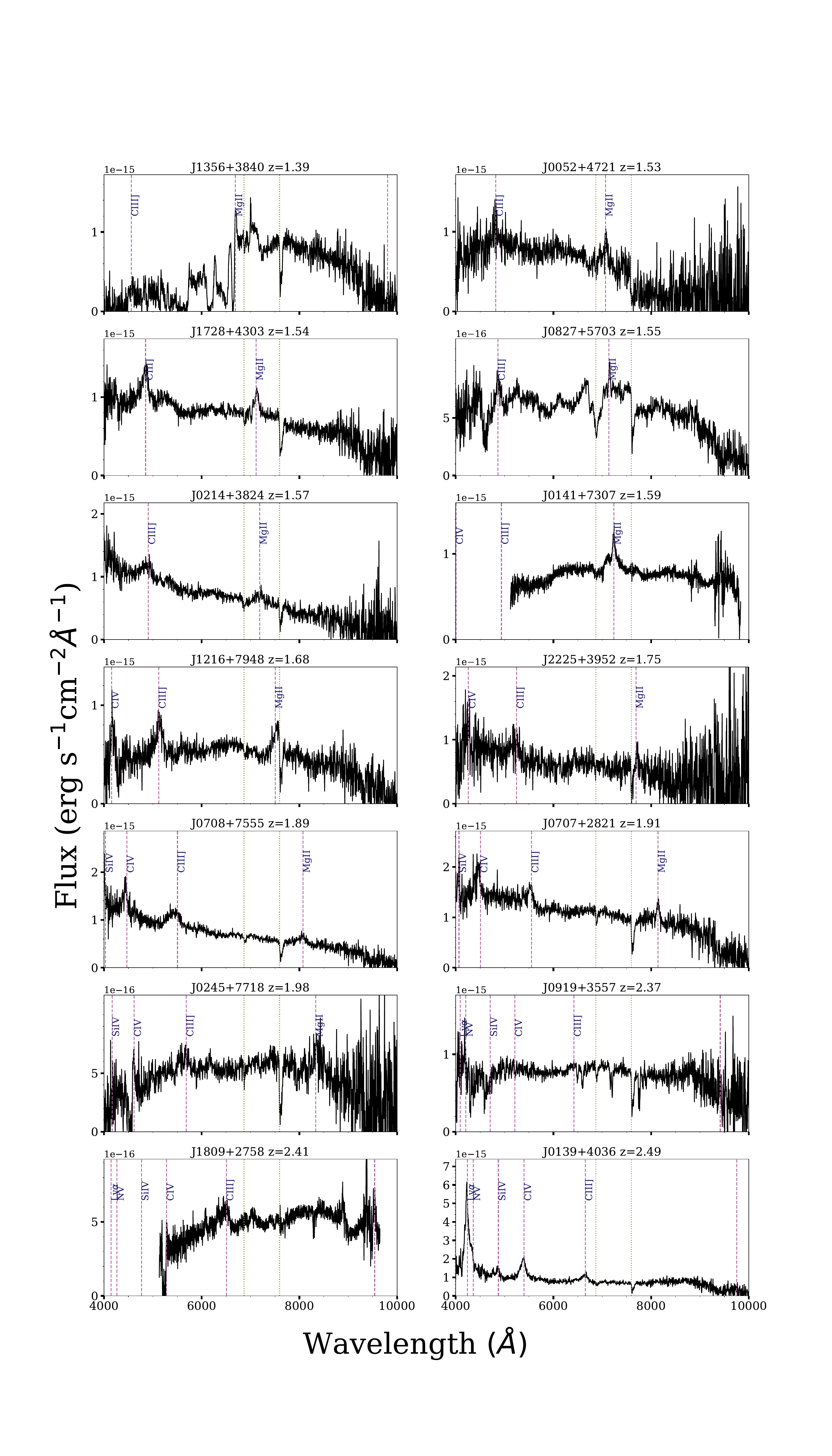}
    \caption{(Continued)}
\end{figure*}

\clearpage

\subsection{Unclassified sources observed in AllBRICQS} \label{sec:unclass}

\begin{figure*}[h]
    \centering
    \includegraphics[width=\textwidth, trim=0cm 0cm 0cm 0cm, clip]{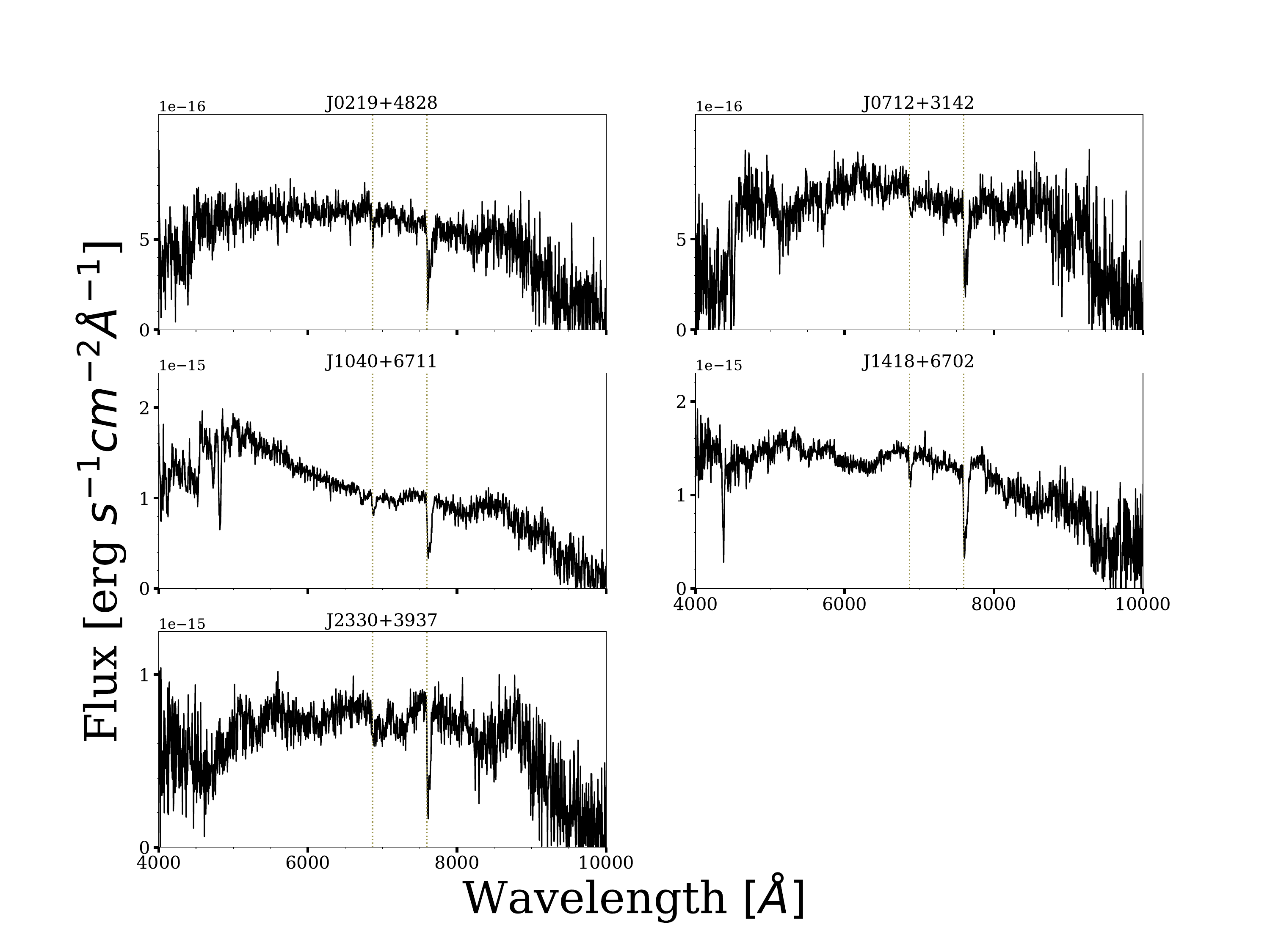}
    \caption{Spectra of the unclassified sources, presented in order of increasing RA. The $y$-axis shows the scaled flux, aligned with contemporaneous broadband photometry. Green dotted vertical lines at 6867.2~\AA\ and 7593.7~\AA\ mark regions affected by telluric absorption, which may impact spectral interpretation.}
    \label{fig:UnknownSED}
\end{figure*}

Figure~\ref{fig:UnknownSED} presents the spectra of the AllBRICQS unclassified sources, arranged in order of ascending RA. Spectra with incomplete wavelength coverage correspond to targets observed with either XLT or LJT, where partial spectral data were obtained.

\setlength{\abovecaptionskip}{0cm}

\clearpage
\subsection{Stars observed in AllBRICQS} \label{sec:stars}

\begin{figure*}[h]
    \centering
    \includegraphics[width=\textwidth, trim=0cm 2cm 0cm 0cm, clip]{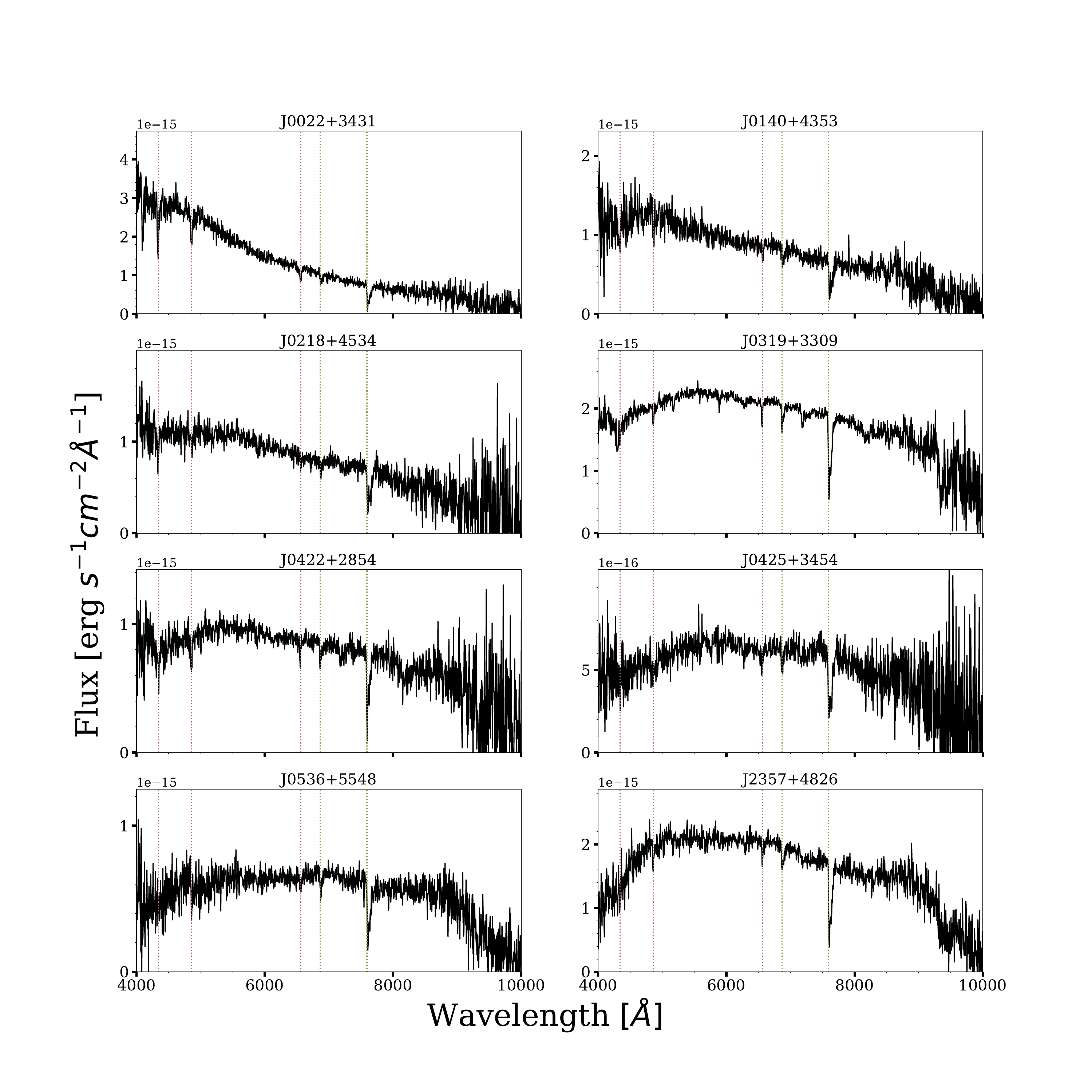}
    \caption{Spectra of the stars, presented in order of increasing RA. The $y$-axis shows the scaled flux, aligned with contemporaneous broadband photometry. Pink dotted vertical lines indicate the positions of Balmer absorption lines. Green dotted vertical lines at 6867.2~\AA\ and 7593.7~\AA\ mark regions affected by telluric absorption, which may impact spectral interpretation.}
    \label{fig:StarSED}
\end{figure*}

Figure~\ref{fig:StarSED} presents the spectra of the AllBRICQS stars, arranged in order of ascending RA. Spectra with incomplete wavelength coverage correspond to targets observed with either XLT or LJT, where partial spectral data were obtained.

\clearpage
\section{Quasars on SDSS Footprint} \label{sec:sdss_img}
\begin{figure*}[h]
    \centering
    \includegraphics[width=0.8\textwidth, keepaspectratio]{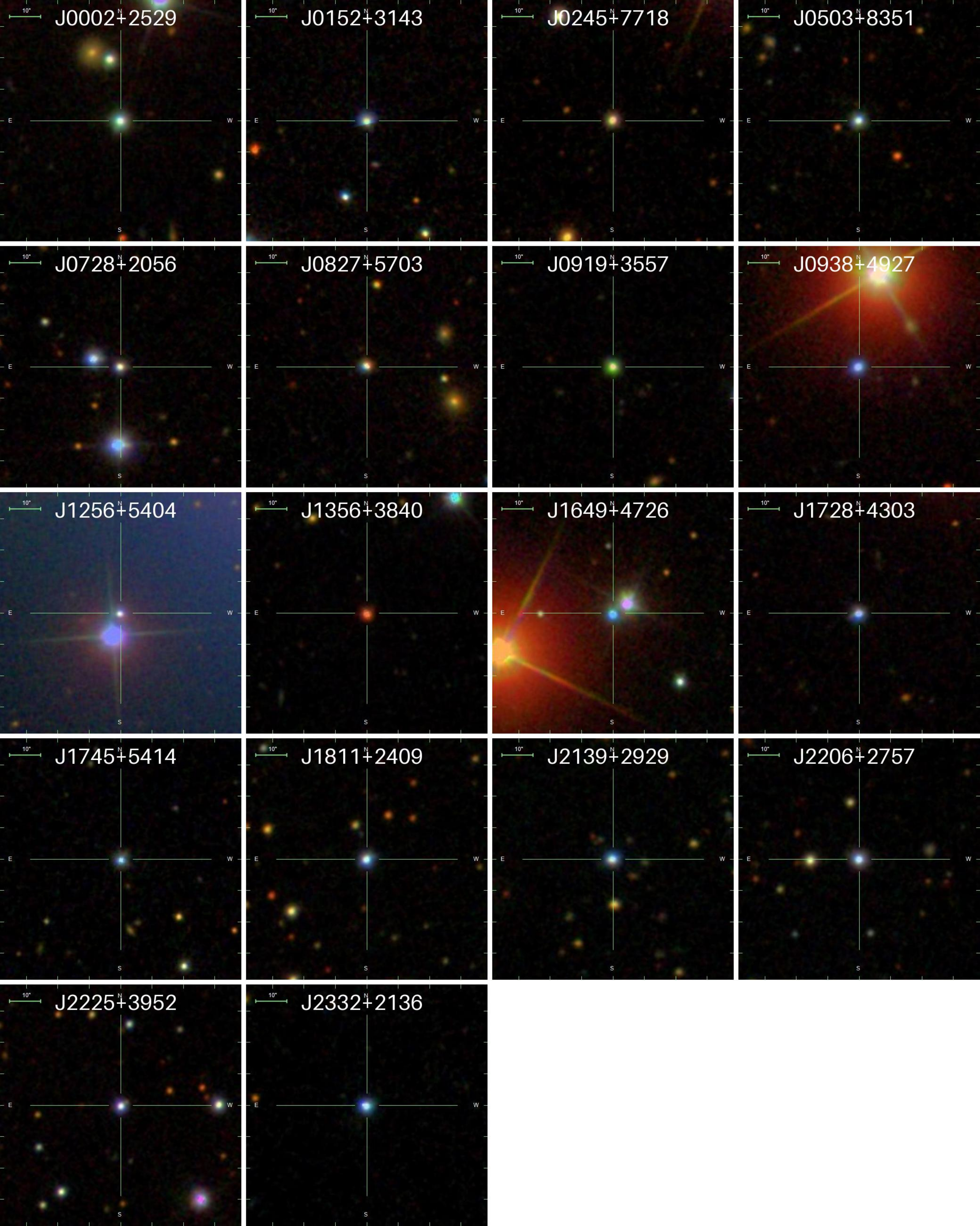}
    \caption{Cutout images of 18 newly discovered quasars that lie within the SDSS footprint but are not included in the SDSS DR16Q catalog \citep{lyke_sloan_2020}. The name of each quasar is displayed at the top of its respective panel. A scale bar indicating 10 arcseconds is shown in the upper left corner of each panel.
    }
    \label{fig:comimg_sdss}
\end{figure*}

Figure~\ref{fig:comimg_sdss} shows the gallery of 18 quasars located within the SDSS footprint \citep{ahumada_16th_2020}.

\setlength{\abovecaptionskip}{0.5cm}

\clearpage
\section{Optical Spectra} \label{sec:example_spec}
An example of the optical spectrum data is shown in Table~\ref{tab:ExampleSpec}, which provides a partial excerpt of the spectrum for J2114+2524 in ASCII format. The full optical spectra of all 62 quasars are available in machine-readable format.

\newcolumntype{C}[1]{>{\centering\arraybackslash}p{#1}} 

\begin{longtable}{
p{1.5cm} C{3.0cm} C{3.0cm} 
}
\caption{Spectrum of J2114+2524} \label{tab:ExampleSpec} \\
\hline
$\lambda$ & $f_{\lambda}$ & $f_{\lambda}$ Uncertainty \\
(\text{\AA}) & (erg s$^{-1}$ cm$^{-2}$ \text{\AA}$^{-1}$) & (erg s$^{-1}$ cm$^{-2}$ \text{\AA}$^{-1}$) \\
\hline
\endfirsthead

\hline
\multicolumn{3}{c}{\textit{Continued from previous page}} \\
\hline
$\lambda$ (\text{\AA}) & $f_{\lambda}$ (erg s$^{-1}$ cm$^{-2}$ \text{\AA}$^{-1}$) & Uncertainty (erg s$^{-1}$ cm$^{-2}$ \text{\AA}$^{-1}$) \\
\hline
\endhead

\hline
\multicolumn{3}{c}{\textit{Continued on next page}} \\
\hline
\endfoot

\hline
\endlastfoot


4002.4 & 1.588E-15 & 2.061E-16 \\
4006.9 & 1.211E-15 & 2.035E-16\\
4011.4 & 2.161E-15 & 2.010E-16\\
4016.0 & 1.153E-15 & 1.901E-16\\
4020.5 & 1.454E-15 & 1.962E-16\\
4025.0 & 9.426E-16 & 1.961E-16\\
4029.5 & 8.905E-16 & 1.857E-16\\
4034.1 & 1.462E-15 & 1.877E-16\\
4038.6 & 2.312E-15 & 1.997E-16\\
4043.1 & 1.464E-15 & 1.878E-16\\

\end{longtable}
\tablecomments{Table \ref{tab:ExampleSpec} is published in its entirety in the machine-readable format. A portion is shown here for guidance regarding its form and content.}

\end{document}